\documentclass{jfm}
\usepackage[utf8]{inputenc}
\usepackage{graphicx}
\usepackage{natbib}
\usepackage{amsmath,amssymb,amsfonts,amssymb}
\usepackage{subfig}
\usepackage{setspace}
\usepackage{float}
\usepackage{microtype}
\usepackage[normalem]{ulem}
\usepackage{ragged2e}
\usepackage{xcolor}
\usepackage{tabularx}

\DeclareCaptionJustification{justified}{\justifying}
\definecolor{fgreen}{rgb}{0.0, 0.5, 0.0}
\definecolor{dblue}{rgb}{0.2, 0.2, 0.6}
\definecolor{springgreen}{rgb}{0.09, 0.45, 0.27}
\definecolor{dartmouthgreen}{rgb}{0.05, 0.5, 0.06}
\definecolor{egyptianblue}{rgb}{0.06, 0.2, 0.65}
\definecolor{fireenginered}{rgb}{0.81, 0.09, 0.13}
\definecolor{forestgreen}{rgb}{0.0, 0.27, 0.13}
\definecolor{harvardcrimson}{rgb}{0.79, 0.0, 0.09}
\definecolor{amaranth}{rgb}{0.9, 0.17, 0.31}

\usepackage{hyperref}
\hypersetup{
	hyperindex,
	breaklinks,
	colorlinks=true,
	linkcolor=blue,
	citecolor=magenta,
	bookmarks=true,
	bookmarksopen=true,
	bookmarksopenlevel=2,
	pdfstartpage={1},
	pdfstartview={FitH},
	pdfview={FitH 0},
	pdfauthor={B. F. Farrell and P. J. Ioannou},
	pdftitle={}}

\usepackage{ifthen}

\def\U{\bm{\mathsf{U}}}

\def\C{\bm{\mathsf{C}}}
\def\I{\bm{\mathsf{I}}}

\def\P{{\bf P}}

\def\A{{\bf A}}

\def\U{\bm{\mathsf{U}}}

\def\C{{\bf C}}

\def\I{{\bf I}}

\def\P{{\bf P}}

\def\C{{\bf C}}

\newcommand\redsout{\bgroup\markoverwith{\textcolor{red}{\rule[0.5ex]{2pt}{0.4pt}}}\ULon}

\newcommand{\be}{\begin{equation}}
\newcommand{\ee}{\end{equation}}
\newcommand{\bdm}{\begin{equation*}}
\newcommand{\edm}{\end{equation*}}
\newcommand{\bea}{\begin{eqnarray}}
\newcommand{\eea}{\end{eqnarray}}

\newcommand{\partialf}[2]
{
 \ifthenelse{\equal{#1}{}}{\frac{\partial}{\partial #2}}{\frac{\partial #1}{\partial #2}}
}

\renewcommand{\(}{\left(}
\renewcommand{\)}{\right)}
\renewcommand{\[}{\left[}
\renewcommand{\]}{\right]}
\newcommand{\<}{\left\langle}
\renewcommand{\>}{\right\rangle}

\newcommand{\df}{\textrm{d}}

\providecommand\bcdot{\boldsymbol{\cdot}}

\newcounter{saveeqn}%

\def\bt{\tilde{\beta}}

\def\st{\sin{\vartheta}}

\def\xv{\mathbf{x}}

\providecommand\bcdot{\boldsymbol{\cdot}}

\newcommand{\ut}{u_\tau}

\renewcommand{\U}{\mathbf{U}}
\renewcommand{\u}{\mathbf{u}}

\shorttitle{POD stucture and dynamics in turbulent Poiseuille flow}
\shortauthor{M-A. Nikolaidis et al}

\title{POD-based study  of turbulent  plane Poiseuille flow: comparing structure and dynamics between  quasi-linear simulations  and DNS}

\author{Marios-Andreas~Nikolaidis\aff{1},
 Petros J. Ioannou\aff{1,2}\corresp{\email{pjioannou@phys.uoa.gr}},
 Brian F. Farrell\aff{2}, Adri\'an~Lozano-Dur\'an\aff{3}
 }
\affiliation{\aff{1}Department of Physics, National and Kapodistrian University of Athens, Athens, Greece
\aff{2}Department of Earth and Planetary Sciences, Harvard University, Cambridge, U.S.A. 
\aff{3}Department of Aeronautics and Astronautics, Massachusetts Institute of Technology, Cambridge, U.S.A.
}

\begin{document}

\maketitle

%
%
%
%

\begin{abstract}
Turbulence in the restricted nonlinear (RNL) dynamics is analyzed and compared with DNS 
of Poiseuille turbulence at $R=1650$. The structures are obtained by POD analysis of the two 
components of the flow partition used in RNL dynamics:  the streamwise-mean flow 
and  fluctuations. POD analysis  of the streamwise-mean flow indicates that 
the dominant POD modes,  in both DNS and RNL, are roll-streaks harmonic in the spanwise. However, we conclude that  
these POD modes do not occur in isolation but rather are  Fourier 
components of a  coherent roll-streak structure.  
POD analysis of the fluctuations in DNS and RNL reveals similar complex structures consisting in part of oblique waves
collocated with the streak. The origin of these structures is identified by their correspondence to POD modes predicted using a 
stochastic turbulence model (STM).  These predicted POD modes are   dominated by the optimally growing structures on the streak, which the 
STM predicts correctly to be of sinuous oblique wave structure.
This close  correspondence between   the roll-streak structure  and the associated fluctuations  in DNS, RNL and the STM implies that
the self-sustaining mechanism operating in DNS is essentially the same as that in RNL, which has been previously   associated with optimal perturbation 
growth on the streak.  
\end{abstract}


\maketitle


\section{Introduction}

The Proper Orthogonal Decomposition (POD) analysis of a time-dependent velocity field proceeds by first obtaining a 
time-mean flow and then forming the average spatial covariance 
of the components of the velocity 
fluctuations about this mean flow.   The eigenfunctions of this covariance are the POD modes of the flow \citep{Lumley-1967,Aubry-etal-1988,Moin-Moser-89,Berkooz-etal-1993,Sirovich-etal-1990,Moehlis-etal-2002,Hellstrom-etal-2011,Hellstrom-Smits-2017}.
 The Eckart-Young-Mirsky theorem \citep{Eckart-Young-36,Mirsky-60} assures that the POD modes constitute
 an optimal basis for  the  fluctuation covariance consistent with which the POD modes comprise an orthogonal 
 basis ordered in contribution to the fluctuation variance,
which can be used to study and compare simulations 
to other simulations or to observations.
POD modes have also been proposed as a means to identify structures appearing in the flow and in particular 
coherent structures.   
However, caution is needed in the use of POD modes for purposes other than the optimally compact 
representation 
of a covariance obtained from a data set, which is the purpose validated by the Eckart-Young-Mirsky 
theorem.  One reason for using caution when interpreting POD modes is that
there is arbitrariness in the structures that produce a given covariance. 
 As \cite{Cantwell-1981} points out, there is no unique relationship between
the covariance obtained from a data set and the states that produced it.  Indeed,   the most general class of 
states that produce the same
covariance is that of  a unitary transformation of the  POD modes \citep{Schroedinger-36,Farrell-Ioannou-2002-perturbation-II}. 
It follows that, while the POD modes provide 
a basis for optimally representing the fluctuation variance, there is no reason to 
expect that  the members of this basis will resemble structures appearing in the flow and
ancillary information is required to connect the POD modes to structure.  
A trivial example of such ancillary information would be a rank one covariance in  which a single 
 POD mode identifies the only structure that appears in the flow.  
 This example is perhaps not as trivial as it seems because often a single POD mode 
 does dominate the variance, as revealed by its eigenvalue being 
 substantially larger than the others, in which case one can expect this dominant 
 mode to be prominently seen in  the flow.  
Unfortunately,  a second structure can not in general be identified with the POD 
mode having the next largest eigenvalue.  This is because, except in the special cases such as the Langevin system 
mentioned in the next paragraph, the structure of the second most prominent contributor to variance appearing 
in the flow will not in general be orthogonal to the first and 
so its structure will be influenced by the requirement that it be 
projected onto the subspace orthogonal to the first, and so on for all other POD modes.
Given that the POD modes form a basis, what is needed  as ancillary information 
to obtain structure identification
is the amplitude and displacement among the POD modes so that their superposition is accounted for in forming the structure.  
From this viewpoint the POD modes are regarded as constituting a compact basis but the structure of the individual POD 
mode is not regarded as providing complete structure information which requires properly accounting for the superposition of the POD modes.
For a  turbulence that arises from a dynamics that is  homogeneous in  a given
coordinate, random perturbations eventually mix any coherent structures that arise in that homogeneous coordinate so that
the POD modes comprise a Fourier basis in that coordinate and structure information is encoded in the 
amplitudes and phases of the harmonics.
If one makes the random phase assumption  then the 
fluctuations have minimal coherence and in that coordinate take  the form of a  spatially stochastic process, while if one makes the zero phase 
assumption, a   compact structure is obtained in which  the harmonics add coherently at the chosen origin.

A case for which explicit interpretation of the POD modes 
can be made is that of a normal  linear dynamical system  of  Langevin form forced white in space and time.
In this case  the POD modes identify 
the eigenmodes of the system and  the  real parts of their eigenvalues \citep{North-84}.  This example  has led to the inference that the POD 
modes can be used to infer information about both structure and dynamics  in turbulent flows.
This inference is misleading 
except under the highly restricting assumptions mentioned. Not only are
the individual POD modes not necessarily structures  that appear in the flow, 
but neither do they provide an optimal basis for the flow dynamics.
In particular, as the dynamics in wall-bounded flows is
non-normal, the  growing structures that give rise to the POD modes are very different from  the POD modes themselves
and  exploiting POD modes to reduce the dimension of the system maintaining the turbulence
requires retaining a separate set of growing structures  in addition to the set of retained POD modes when forming the basis supporting the dynamics
\citep{Farrell-Ioannou-1993e,Farrell-Ioannou-2001-accurate,Rowley-2005}.

POD analysis   was originally advanced  as a 
method for identifying coherent structures 
in wall turbulence and investigating their dynamics  \citep{Lumley-1967,Berkooz-etal-1993}.
It was presumed that as the coherent structures represent a substantial fraction of the variance, the POD modes  would 
identify these structures.
However, this project of associating the POD modes with coherent structures faced  the difficulties mentioned above
and in addition  issues specific to the wall-turbulence problem.
The model problems addressed in studies of wall-turbulence are  homogeneous in the streamwise and spanwise direction. 
Consistent with the above discussion, the structures in these turbulent flows explore all spanwise and streamwise locations equally,  resulting in 
a time mean flow and covariance  that are
asymptotically   homogeneous  in the spanwise and streamwise directions. The mean flow then depends  only on the cross-stream direction 
and the  covariance depends only on the relative separation of points in the spanwise and streamwise directions. This implies that the 
POD modes are harmonic in the 
spanwise and streamwise coordinates and eigenalysis of the covariance  
can only identify the time mean variance of these harmonic POD modes together with their associated
cross-stream structure, the cross-stream being the only inhomogeneous direction,   
but  leaves their amplitude, phase, and therefore their
structure in the spanwise and streamwise directions undetermined.  
This  absence of information about the amplitude and phase of the POD modes in the spanwise and streamwise 
renders the POD modes  incapable of identifying coherent structures 
which, ironically, was the original motivation for studying them.
This was recognized  by \cite{Lumley-1981} who proposed to obtain relative phase information in the homogeneous coordinates
from higher order statistics 
and in this way to complete the identification of the coherent structures using POD analysis. 
In the pursuit of this goal, of particular interest are the results and  methods of \cite{Moin-Moser-89}
who used  statistical methods for estimating the POD modes in a turbulent channel flow by which they identified a dominant
coherent structure consisting of a compact streamwise elongated low-speed streak flanked by a pair of compact rolls,
which they associated with the coherent structure that arises in the bursting process; \cite{Jimenez-2018} contains a recent review of these methods.
The result of these attempts to obtain the phases of the POD modes 
by statistical means is to elicit structure similar to that predicted 
by the minimum entropy assumption of aligning the phases among
the modes to produce a maximally compact structure. 
A related  problem  of identifying  traveling
coherent  structures using POD analysis was addressed  using slicing and centering methods 
\citep{Rowley-2000,Cvitanovic-2012,Cvitanovic-etal-2013}. An analogous procedure is employed in this paper  to isolate
the low-speed streak and its associated fluctuation field.

We have reviewed the conceptual basis for, as well as the limitations of,
 POD-based analyses.  
Our objective in this work is to adapt POD-based analysis to facilitate the study of aspects of the dynamical
mechanism supporting the turbulence. We do this by 
applying modified  POD analysis methods to compare structure and dynamics
between the restricted nonlinear (RNL) quasi-linear system 
and the associated DNS.
The motivation for doing this comparison is that  RNL 
is obtained directly from the Navier-Stokes equations with only the omission 
of the nonlinear interaction of the streamwise varying  components of the flow. 
This elimination of the nonlinear perturbation term greatly simplifies the turbulence dynamics while 
retaining the essential mechanisms supporting turbulence which facilitates study of these mechanisms.
Important for our study is that
RNL sustains
a realistic turbulence despite its highly simplified dynamics. Also, as we will describe further, 
the RNL system is to a substantial degree analytically characterized
\citep{Farrell-etal-2016-VLSM}.
It follows that if a convincing case can be made for essential similarity 
in the structure, and by extension the dynamics underlying turbulence in DNS and RNL,  then the simplicity of the dynamics of
RNL turbulence can be exploited to provide insight into the mechanism of wall-turbulence. 

We proceed by briefly reviewing the formulation of RNL as a quasi-linear approximation
of the Navier-Stokes (N-S) equations, the simplifications that result from this 
approximation,  and the insights  this approximation provides 
for understanding the mechanism of wall-turbulence \citep{Farrell-etal-2016-PTRSA}.
To obtain the  RNL approximation, the N-S equations are first decomposed into equations governing the 
streamwise-mean flow and the fluctuations from the streamwise-mean. 
At this point no approximations have been made to the Navier-Stokes equations.
The RNL approximation consists in neglecting  the fluctuation-fluctuation interactions in the fluctuation equations.
It follows that RNL dynamics 
comprises the quasi-linear interaction between the time-dependent streamwise-mean flow and the fluctuation
field. 
It is important to recognize that the fluctuation equation,
which has been isolated from the nonlinear streamwise mean equation by 
this partition of the dynamics,  is linear in the fluctuations. 
However, while the fluctuation equation is linear in the fluctuations, it is also time-dependent due to the time dependence of 
the streamwise mean flow, and therefore it can extract energy from the mean flow through the parametric mechanism, which is supported
by  concatenation of  non-normal 
growth events.
This time-dependent parametric interaction with the 
mean flow   provides periods of fluctuation
growth and decay. Given that the fluctuation field is 
bounded, its time-mean growth must be exactly zero or equivalently the 
top Lyapunov exponent of fluctuations growing on the time-varying  
streamwise-mean flow  must be  exactly the real number zero, which requires that
the time-varying mean flow  be regulated to neutral Lyapunov stability by the Reynolds stresses of the fluctuations.
This implies that the fluctuation field of RNL turbulence lies in  the subspace of the 
Lyapunov vectors   of the time-varying streamwise mean flow that have zero Lyapunov exponent
and  the mean flow is regulated  by feedback from the Reynolds stresses of these Lyapunov vectors to neutral Lyapunov stability.  
 This simplification of  the turbulence to a subset of analytically characterized fluctuations supported by as few as a single 
 streamwise-varying harmonic
 occurs spontaneously in the RNL system.
The fact that RNL is supported on the  small set of  Lyapunov vectors with precisely zero Lyapunov exponent and
that the time mean state is feedback regulated  to exact Lyapunov neutrality provides comprehensive analytic characterization of 
both the fluctuations and the regulation of the statistical mean state of RNL turbulence.

 It is interesting to note that this quasilinear adjustment to neutral stability constitutes a 
 solution for the statistical state of the turbulence to second order that 
 vindicates the program of \cite{Malkus-1956} to obtain a 
quasi-linear equilibration  identifying the statistical mean state of turbulence in 
shear flow -  it only being required to recognize
that it is not the inflectional or the viscous instability of the time-mean turbulent profile 
that is neutralized, as proposed by \cite{Malkus-1956} and critiqued by \cite{Reynolds-Tiederman-1967}, but
rather the parametric instability of the
streamwise mean flow, which  arises from the temporal variation of the roll-streak structure (R-S). 
An important  insight, inherent in the RNL formulation, arises from isolation in
the mean equation of the primary coherent structure, which is the R-S.
This partition of structure in the turbulence into mean and 
fluctuation with the R-S
spontaneously forming in the mean equation poses a fundamental constraint on
mechanistic theories of wall-turbulence. The necessary inference is that the R-S in RNL is
maintained by 
fluctuation Reynolds stresses arising from a non-normal parametric growth
process and not by a 
fluctuation-fluctuation scattering regeneration  as proposed by \cite{Trefethen-etal-1993,Farrell-Ioannou-1993e,Gebhardt-Grossmann-1994}. 
In these regeneration mechanisms,  optimal  perturbations that have been recycled from  turbulent debris,
typically ascribed in the case of wall-turbulence to streak breakdown, through their growth  sustain the turbulence \citep{Jimenez-Moin-1991,Jimenez-2018}.  
 Another example of the regeneration mechanism sustaining turbulence is the
baroclinic turbulence of the midlatitude atmosphere \citep{DelSole-2007,Farrell-Ioannou-2009-closure}. It was recently shown 
by \cite{Lozano-Duran-etal-2021} that non-normal amplification of fluctuations 
regenerated through fluctuation-fluctuation interactions in an externally maintained  stable time-independent mean flow can sustain
a turbulent fluctuation field.
Clearly,  the nonlinear scattering regeneration mechanism is available to support turbulence. 
 However, RNL 
turbulence makes a radical departure from this regeneration mechanism by sustaining turbulence without any  fluctuation-fluctuation nonlinearity. 
The mechanism sustaining turbulence in RNL is
consistent conceptually with the  SSP mechanism advanced in   \cite{Hamilton-etal-1995} and illustrated by the toy model based 
studies of \cite{Waleffe-1997},  in that the streak in 
RNL  is similarly supported by roll-induced  lift-up with the roll in turn being maintained  by 
torques from 
Reynolds stresses produced by the associated fluctuation field.
However, in understanding the SSP mechanism the origin, maintenance, and collocation with the streak of the roll-inducing torques is the central dynamical problem.  
The primary mechanisms advanced to account for the roll-inducing torques are Reynolds stresses arising 
from modal structures  \citep{Waleffe-1997,Waleffe-2001,Hall-Sherwin-2010}
 and Reynolds stresses arising from optimally growing transient structures \citep{Schoppa-Hussain-2002,Farrell-Ioannou-2012}.  
This question of the mechanism underlying the SSP has been addressed in
recent work which verified  that turbulence maintenance 
is essentially unaffected when modal instability is suppressed at every 
time step in  a DNS of constant mass-flux pipe-flow or in RNL simulations of Couette turbulence,  which provides constructive proof 
that instability is not related to  turbulence, at least in these systems \citep{Farrell-Ioannou-2012,Lozano-Duran-etal-2021}.
In this work we provide evidence that in both DNS and RNL the  Reynolds stress that induce torques maintaining the 
SSP arise from transiently growing structures.  However, in RNL the non-normal fluctuations arise from 
parametric interaction with the mean flow rather than from fluctuation-fluctution nonlinearity.  Regardless of how
the SSP is maintained,
the SSP  mechanism for sustaining the R-S is fundamentally 
different from the mechanism proposed by e.g. \cite{Jimenez-2013lin,Jimenez-2013,Jimenez-2022}
in which streaks arise as   scars left in the streamwise velocity 
from the linear growth of  episodically   excited optimal perturbations \citep{Encinar-Jimenez-2020}.


The goal of this work is to use POD-based diagnostics to demonstrate that 
DNS and RNL turbulence exhibit compellingly similar dynamical structures, which suggests similarity in the dynamics underlying
these structures. This dynamics is maintenance of streak-collocated roll circulations by Reynolds stress torques
arising from the transient growth of optimal perturbations. This is the universal mechanism 
 by which optimal perturbations  induce streak-collocated roll circulations \citep{Farrell-Ioannou-2012,Farrell-Ioannou-2017-bifur,Farrell-Ioannou-2022}.
 This mechanism will be also  the subject of a companion paper focusing on the dynamics of R-S  formation using  the same DNS and RNL dataset \citep{Nikolaidis-RS-2023}.

%
%

In this paper we first compare  the POD modes 
of the streamwise mean flow  predicted under the assumption of 
spanwise homogeneity in DNS and RNL.  
We then obtain a converged estimate, in both DNS and RNL, of the coherent R-S by collocating the observed streaks in both systems.
Having obtained the coherent R-S in DNS and RNL, we
compare their structures, which are found to be remarkably similar. 
Having a converged estimate of the R-S structures in these systems, we next  verify  that the 
POD amplitudes obtained under the assumption of spanwise homogeneity are convincingly coincident with the Fourier amplitudes 
arising from Fourier analysis of these  R-S structures. 
In this way, we verify that the collocation process  correctly identifies the phases of the POD modes.    
Together these results verify the spontaneous symmetry breaking in the 
spanwise direction by the emergence of the R-S instability, as predicted by the S3T SSD
stability analysis, is occurring in both DNS and RNL \citep{Farrell-Ioannou-2012,Farrell-Ioannou-2017-bifur}.  

Having obtained the time-mean
R-S structure in both DNS and RNL we
turn next to obtaining POD analyses of the streamwise inhomogeneous fluctuations 
about the mean R-S in both DNS and RNL and verify that the fluctuation fields educed by this fluctuation component   POD analyses are consistent
with the prediction of oblique waves as being responsible for maintaining the
 coherent streamwise roll in the SSP \citep{Farrell-Ioannou-2022}. 
The fluctuation POD modes  are then shown to be consistent with predictions for optimally growing structures over 
typical temporal correlation times in these turbulent flows by comparing them with
the average structure of stochastically excited  evolving  fluctuations  over 30 advective time units.
This identification of wave-like structures  maintaining 
the R-S in DNS and RNL with optimally growing perturbations constitutes 
a compelling argument that the turbulence in both systems is supported by the SSP 
that has been analytically identified in RNL dynamics  and that this SSP is maintained by Reynolds stress torques produced by optimally growing perturbations.

\section{DNS and its RNL approximation}\label{sec:reduction}

We  study a pressure driven constant mass-flux plane Poiseuille flow in a 
channel which is doubly periodic  in the streamwise, $x$, and
spanwise, $z$, direction.
The incompressible non-dimensional Navier-Stokes equations governing
the channel flow are decomposed
into equations for the streamwise mean flow, $\U=(U,V,W)$, and the  fluctuations, $\u=(u,v,w)$,
as follows:
\begin{subequations}
\label{eq:DNS}
\begin{align}
\partial_t\U&+ \U \bcdot \nabla \U  - G(t) \hat{\mathbf{x}} + \nabla P - R^{-1} \Delta \U = - \overline{\u \bcdot \nabla \u}~,
\label{eq:NSm}\\
 \partial_t\u&+   \U \bcdot \nabla \u +
\u \bcdot \nabla \U  + \nabla p-  R^{-1} \Delta  \u
= -(\u \bcdot \nabla \u - \overline{\u \bcdot \nabla \u})~.
 \label{eq:NSp}\\
&\nabla \bcdot \mathbf{U} = 0~,~~~\nabla \bcdot \mathbf{u} = 0~.\label{eq:NSdiv0}
\end{align}\label{eq:NSE0}\end{subequations}   
 No-slip impermeable boundaries are placed at $y=0$ and $y= 2$, in the wall-normal variable. The  
 pressure gradient $G(t) \hat{\mathbf{x}}$ is adjusted in time to maintain  constant mass flux, $\hat{\xv}$ is the unit vector in the streamwise direction. An  
overline, e.g. $\overline{\u\bcdot \nabla \u}$,  denotes averaging  in $x$. Capital letters indicate  streamwise averaged quantities.
Lengths have been made non-dimensional by  $h$, 
the channel's half-width, velocities by $\langle U \rangle_c$, the center velocity of the time-mean flow, and time by $h/\langle U\rangle_c$. The Reynolds number is $R= \langle U \rangle_c h / \nu$, with  $\nu$ the kinematic viscosity.

The  corresponding  RNL equations  are obtained by suppressing nonlinear interactions among streamwise-varying flow components 
in the fluctuation equations  resulting in the right hand side of \eqref{eq:NSp} being neglected. The RNL equations are:
 \begin{subequations}
\label{eq:QL}
\begin{align}
\partial_t\U&+ \U \bcdot \nabla \U  - G(t) \hat{\mathbf{x}} + \nabla P - R^{-1}\Delta \U = - \overline{\u \bcdot \nabla \u}~,
\label{eq:QLm}\\
 \partial_t\u&+   \U \bcdot \nabla \u+
\u \bcdot \nabla \U  + \nabla p-  R^{-1} \Delta  \u
= 0~. 
 \label{eq:QLp}\\
&\nabla \bcdot \mathbf{U} = 0~,~~~\nabla \bcdot \mathbf{u} = 0~.\label{eq:QLdiv0}
\end{align}\label{eq:QLE0}\end{subequations}   
Under this quasi-linear restriction, the fluctuation field interacts nonlinearly  only with the mean, $\U$, flow and  not with itself. This quasi-linear
restriction of the dynamics results in
the spontaneous collapse of the support of the fluctuation dynamics to a 
small subset of streamwise Fourier components.  
It is important to recognize that this restriction in the support of  RNL turbulence to a small 
subset of streamwise Fourier components is not imposed but rather is a property of the dynamics with significant implication.
The components that are retained spontaneously by 
the RNL dynamics   identify the streamwise harmonics that are dynamically active
in the sense that this subset of streamwise harmonics participate  in
the parametric instability that sustains the fluctuation component of the  turbulent state  \citep{Farrell-Ioannou-2012,Thomas-etal-2014,Thomas-etal-2015,Farrell-Ioannou-2017-sync,Nikolaidis-Ioannou-2022}.

\begin{figure*}
\centering\includegraphics[width=26pc]{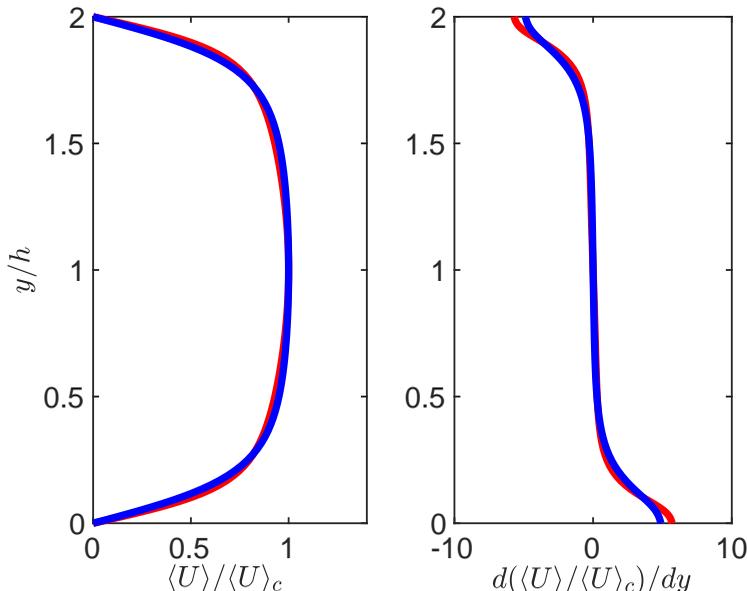}
\vspace{-1em}
\caption{Left panel: The mean velocity profile of the  DNS (red)  and RNL simulations (blue)  
normalized to the average centerline velocity $\langle U \rangle_c$. Right panel: The corresponding normalized mean shear in the two simulations. } \label{fig:Mean_flow_Shear}
\end{figure*}

\begin{figure*}
\centering\includegraphics[width=32pc]{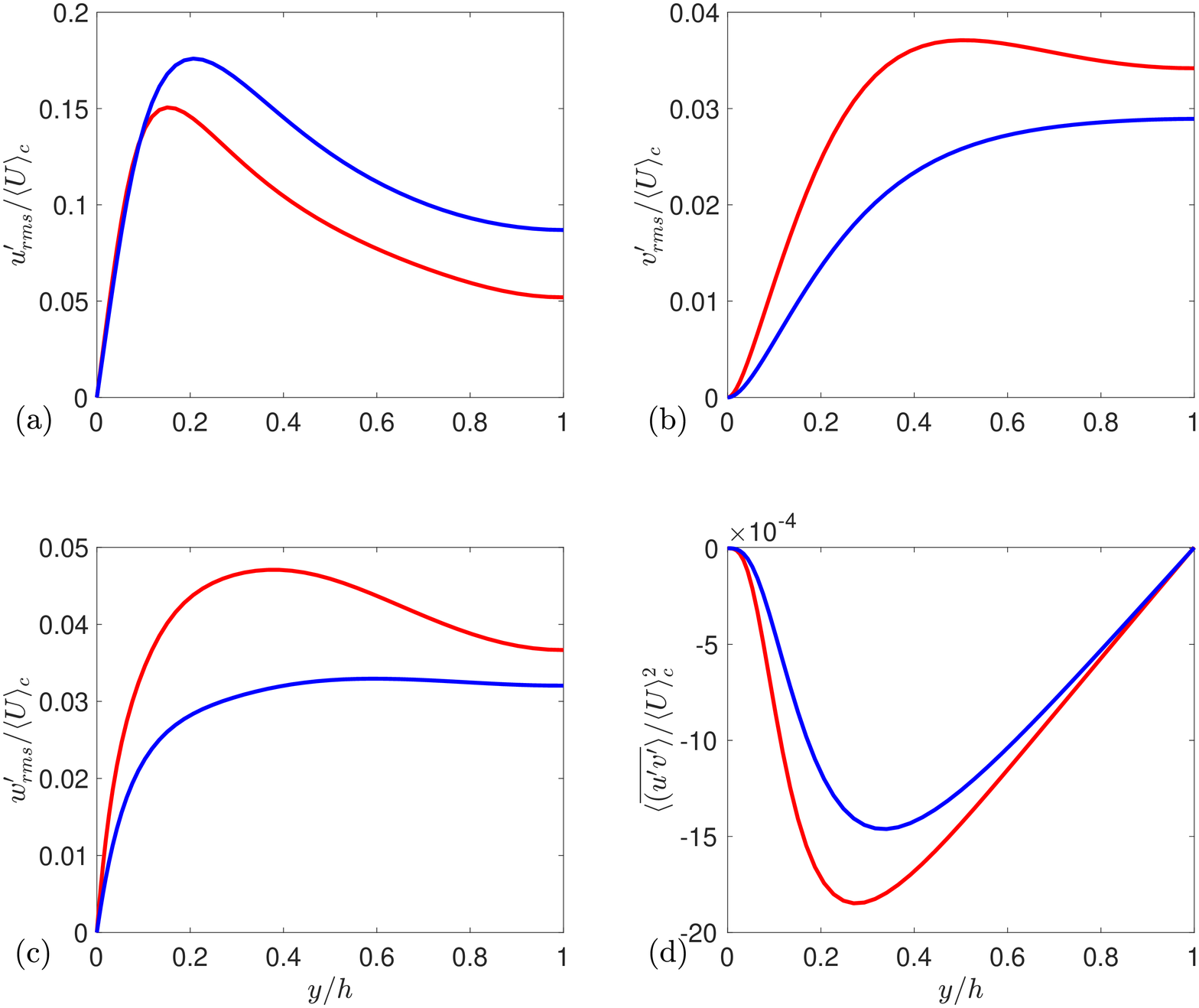}
\vspace{-1em}
\caption{Wall-normal profiles of the rms  of velocity fluctuations (a,b,c) and the tangential Reynolds stress (d) for the DNS (red) and RNL (blue) simulations. } \label{fig:uvw_rms}
\end{figure*}

The data were obtained from a DNS of Eq. ~\eqref{eq:DNS} and from simulation of the associated RNL governed by Eq.~\eqref{eq:QL}. 
The Reynolds number  $R = \langle U\rangle_c h/\nu  =1650$ is 
imposed in both the DNS and  the RNL simulations.
A summary of the parameters of the simulations is given in Table~\ref{table:geometry}.
The time averaged streamwise-mean flow $\langle\U \rangle=\langle U \rangle \hat{\xv}$ and its associated shear   in the DNS and RNL simulation are shown  
in Fig.~\ref{fig:Mean_flow_Shear} and the time-averaged rms profiles of the fluctuations from the mean flow $\langle \U \rangle$, $\u'=\u-\langle U \rangle \hat{\xv}$, are shown in Fig.~\ref{fig:uvw_rms}
The RNL simulation reported here is supported by only three streamwise components with wavelengths $\lambda_x/h = 4 \pi, 2 \pi, 4\pi/3 $,
which correspond to the   three largest  streamwise Fourier components of the channel, $n_x=1,2,3$.
These streamwise Fourier components  sustained in  RNL are not imposed but rather
the RNL spontaneously selects the streamwise Fourier components that are
retained in the turbulent state. We have included 16 streamwise wavenumbers
in the integration of the RNL 
in  order to allow  freedom for the RNL system to select the streamwise wavenumbers that it sustains. 

For the  numerical integration the dynamics were expressed in the form of
evolution equations for the wall-normal vorticity and the Laplacian of
the wall-normal velocity, with spatial discretization and Fourier
dealiasing in the two wall-parallel directions and Chebychev polynomials
in the wall-normal direction~\citep{Kim-etal-1987}. Time stepping was
implemented using the third-order semi-implicit Runge-Kutta method.

\begin{table}
\begin{center}

\begin{tabular}{@{}*{6}{c}}
\break
 Abbreviation   & $[L_x,L_y,L_z]$ & $[L^+_x,L^+_y,L^+_z]$ &$N_x\times N_z\times N_y$& $R_\tau$& $R$ \\
 NL100   & $[4\pi\;,\;2\;,\;\pi]$ &[1264\;,\;201\;,316] &$128\times 63\times 97$&$100.59$&1650   \\
 RNL100 & $[4\pi\;,\;2\;,\;\pi]$ &[1171\;,\;186\;,293] &$ 16\times 63\times 97$&$93.18$&1650   \\
\end{tabular}
\caption{\label{table:geometry}Simulation parameters. $[L_x,L_y,L_z]/h$, where h is the channel half-width, is the domain size in the streamwise, wall-normal and spanwise direction. Similarly,  $[L^+_x,L^+_y,L^+_z]$, indicates the domain size in wall-units. $N_x$, $N_z$ are the number of Fourier components after dealiasing and $N_y$ is the number of Chebyshev components. $R_\tau= \ut h / \nu$ 
is the Reynolds number of the simulation based on the friction velocity  $\ut= \sqrt{ \nu \left.\df \langle U\rangle/\df y\right|_{\rm w}}$,where $\left.\df \langle U\rangle /\df y\right|_{\rm w}$ is the shear at the wall.}
\end{center}
\end{table}

\section{Analysis method used in  obtaining the POD modes }

POD analysis requires 
the two-point same-time spatial covariance of the flow variables. 
The perspective on turbulence dynamics adopted in this work is 
that of the S3T  statistical state dynamics (SSD)  closed at second order \citep{Farrell-Ioannou-2012} and its RNL approximation
\citep{Thomas-etal-2014}.  
The insights on turbulence dynamics obtained by using this SSD proceed from its formulation, which is based on 
using the streamwise-mean and associated fluctuations to express the 
dynamics.  
The choice of the streamwise-mean in the cumulant expansion of this SSD serves to isolate
the dynamics of the dominant coherent structure supporting turbulence, which is the R-S, in the mean equation. 
If the R-S were not supported by the Reynolds stress torque mechanism it would appear in the fluctuation equation.
In order to further isolate the R-S structure  the $k_x=0$ POD analysis is confined to 
deviations of the streamwise mean flow from its spanwise mean.
Adopting the notation $\< \;\boldsymbol{\cdot}\; \>$ for the time average , $\[ \;\boldsymbol{\cdot}\; \]$ for the spanwise average, 
and~$\( \;\boldsymbol{\cdot}\; \)^T$ 
for transposition, the covariance of deviations of the streamwise-mean velocity field from its spanwise-mean is:
\be
C = \< {\cal U} {\cal U}^T \> ~,
\ee
in which
\be
{\cal U}= [U_s  , V_s , W_s  ]^T~,
\ee
is the column vector comprising  deviations of the  three streamwise mean  components 
from their
spanwise-mean, ($[U](y,t),[V](y,t),[W](y,t)$), i.e. $U_s= U-[U]$, $V_s= V - [V]$, and $W_s=W-[W]$.
A requirement for $\cal C$ to be  a covariance is that $\langle U_s \rangle =0 $, $\langle V_s \rangle =0 $ and $\langle W_s \rangle =0 $,
which demands that   $\langle [ U ] \rangle = \langle U \rangle$, $\langle [ V ] \rangle = \langle V \rangle$
and  $\langle [ W ] \rangle = \langle W \rangle$. This condition places a requirement of homogeneity on
the velocity components in $z$, which was verified.

%

The dominant structures in the  fluctuation field are  localized about
 the streamwise streak.  In order to isolate these structures 
the fluctuations are obtained by first collocating the dominant streak 
together with its associated fluctuation field
in the flow 
prior to 
extracting the fluctuations from the dominant streak.  
These fluctuations are used to form the
covariance on which the POD analysis 
of fluctuations from the streamwise mean R-S structure is done,  
as described in section \ref{sec:5}. 
 The covariance of the fluctuation flow field is 
 expressed as:
\be
c = \< \cal{U}'  \cal{U}'^T \> ,
\ee
with
\be
{\cal{ U}}'= [ u, v, w    ]^T,
\ee
which are the column  vector of the  three velocity components of the streamwise varying flow, i.e. the components of   the  velocity deviations
from the average streak structure in the flow.

The POD modes for the mean flow fluctuations and for the fluctuations
from the dominant streak are
obtained by eigenanalysis of the two-point covariances, $C$ and $c$.
The   resulting orthonormal set of eigenvectors is then ordered descending in eigenvalue.
The eigenvalue of each POD mode is its time-averaged contribution  to the  variance of the 
velocity.

To obtain a sufficiently converged covariance to identify the primary POD modes
for the streamwise mean flow requires a long time series of the turbulent flow field. 
Convergence is further facilitated by  assuming the statistical symmetries of the flow fields:
homogeneity in the $x$ and $z$ direction, mirror symmetry in $y$ about the $x-z$ plane at  the channel center,
and mirror symmetry in $z$ about the $y-x$ plane at the channel center.  Because of the mirror symmetry in $y$ the POD modes come in symmetric and antisymmetric pairs
about the $y-x$ plane at the channel center.
Details of the implementation of these symmetries in calculating  the 
POD modes are given in Appendix A. 
Because the roll-streak structure appears at the upper  and lower boundary randomly in spanwise position and time,
 it is  appropriate  to concentrate our analysis on the roll-streak at a single boundary.
The  modes appropriate for composing the roll-streak at the lower boundary are  obtained by adding
the symmetric and antisymmetric in $y$ components of the  POD mode pairs. 
%
%

Statistics of flow quantities   have been verified 
to approach asymptotically 
these symmetries,
as the averaging
time increases.
These statistical symmetries are  not necessary consequences  of  the translation and mirror symmetry
of the NS equations in a periodic channel because the turbulent flow field  may undergo symmetry breaking.
For example, stability analysis of the S3T SSD  of wall-bounded flows in periodic domains
predicts symmetry breaking of spanwise homogeneity  before the turbulent state is established, and an  imperfect  manifestation of this symmetry breaking is 
clearly seen in the related DNS \citep{Farrell-Ioannou-2012,Farrell-Ioannou-2017-bifur}.
Casting the Navier-Stokes equations in 
SSD form permits identification of the instability underlying 
this symmetry breaking, an instability without counterpart in the Navier-Stokes 
equations written in traditional velocity-pressure  component state variables \citep{Farrell-Ioannou-2019-book}.

If an underlying symmetry breaking instability exists in a turbulent system but stochastic 
fluctuations cause the equilibrated modes of this instability  to random walk in 
a homogeneous coordinate so that in the limit of long time the phase 
information localizing the mode is lost, rendering the phase random, then one 
approach to identifying the symmetry breaking mode in a simulation is to obtain an 
approximation to the covariance over short enough times that 
the phase randomization is not complete
while another is to collocate the symmetry breaking structures in the flow so the 
effect of the random walk is removed - we employ the latter method in this work, which is equivalent  to the centering
or slicing method for unveiling coherent structure in data in dynamics with  continuous  symmetries \citep{Rowley-2000,Cvitanovic-2012}.

\section{POD modes of the DNS and RNL streamwise mean flow} \label{sec:stats}

The POD basis for the $k_x=0$ component of the
deviations from the time and spanwise mean velocity
in the DNS and the RNL simulation will first be described under the assumption of spanwise statistical homogeneity . 
Accepting the assumption of statistical homogeneity in  $z$ implies
that the eigenvectors of $C$, which are the POD modes, are single Fourier harmonics in the spanwise
direction  \citep{Berkooz-etal-1993}. Under this assumption the POD  mode, $n_z$,  with
 spanwise  wavenumber $k_z=2 \pi n_z/L_z$ is given by:
\begin{eqnarray}
\Phi_{k_z} = \left(
\begin{array}{c}
  A_{k_z}(y)\\
   B_{k_z} (y)\\
   \Gamma_{k_z}(y)
\end{array}
\right)
e^{i k_z z}~,
\label{eq:PODmodes}
\end{eqnarray}
where $A_{k_z}(y)$ is the streamwise component of the velocity field associated with the POD, $B_{k_z}(y)$ the wall-normal  and $\Gamma_{k_z}(y)$ 
the spanwise component. 
All these components are specified as $N_y$ dimensional column vectors, with  $N_y$ the  number of discretization points in $y$.
At each sampling time     the $3 N_y$ column vector of a $k_z \ne 0$ Fourier component, ${\cal U}_{k_z}(t)$, of the flow field $\cal{U}$   
is obtained and used to form  $N_{k_z}$ average covariances:
\begin{equation}
C_{k_z} = \< {\cal U}_{k_z}(t){\cal U}_{k_z}^\dagger(t) \> ~,\label{eq:Ckz}
\end{equation}
where $N_{k_z}$ is the number of $k_z\ne 0$ Fourier components retained in the simulation and $\dagger$ is the Hermitian transpose.
Eigenanalysis of these covariances determines $3N_y \times N_{k_z}$  eigenvectors  comprising the POD orthonormal basis of the  $k_x=0$ flow field
taking into account the restriction to deviations from the spanwise 
mean mentioned above.  
These POD modes are ordered decreasing in eigenvalue, which 
corresponds to variance.

\begin{figure*}
\centering
\includegraphics[width=30pc]{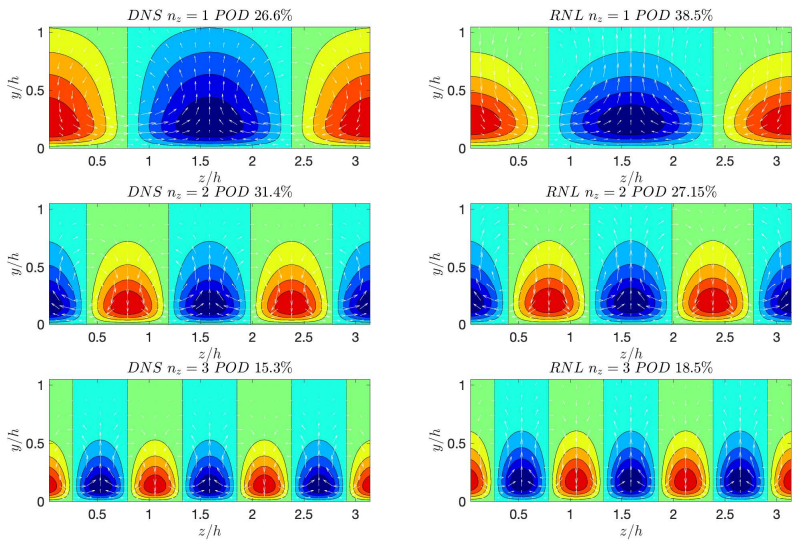}
\vspace{-1em}
\caption{Left panel: The structure of the first 3  POD modes of the streamwise-mean flow 
 appropriate for  the lower boundary obtained  from a  $310000$ advective time units  DNS.
 Right panel: The corresponding modes of the streamwise-mean flow from a $83000$ advective time units  simulation of RNL.
 The contours show levels of the streamwise $U_s$ velocity and the arrows
show the cross stream-spanwise velocity vector $(V_s,W_s)$. The ratio $U_s/V_s$ is in all cases about 10.
Notice that in DNS the POD mode with the largest contribution to the variance is the $n_z=2$ mode,
while in RNL it is the $n_z=1$ mode. The contour level is $0.2$ in all panels.}
 \label{fig:DNS_RNL}
\end{figure*}

\begin{figure*}
\centering\includegraphics[width=26pc]{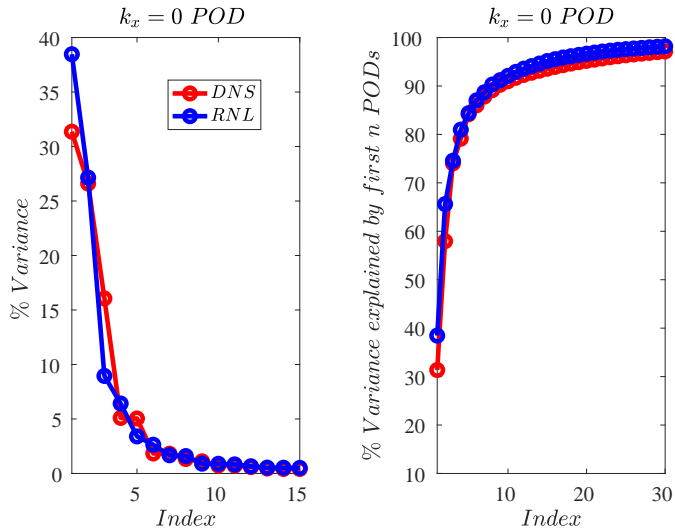}
\vspace{2em}
\caption{Left panel: Percentage variance  of the streamwise-mean ($k_x= 0$) flow  explained by  the POD 
modes  in the DNS and RNL simulation as a function of mode index.
Right panel:
The cumulative variance accounted for by the POD modes in the DNS and RNL simulation as a function of 
the number of POD modes included in the sum.
In DNS the first POD mode has spanwise wavenumber $n_z=2$, the second POD mode  has $n_z=1$. In RNL the first POD mode has spanwise wavenumber $n_z=1$ and the second POD mode  has $n_z=2$.} \label{fig:DNS_RNL_var_270_kx0}
\end{figure*}

As discussed above, because of the statistical homogeneity of the flow in
the $z$ direction,  the $k_x=0$  POD modes come in $\sin(k_z z)$ and $\cos(k_z z)$
pairs, and these  modes contribute equally to the variance.  
 The first three spanwise harmonics of the POD modes of both the DNS and RNL simulation account for $75 \%$ of the  $k_x=0$ variance are shown in Fig. \ref{fig:DNS_RNL}.
 Shown are  both the streak  velocity and  vectors of the corresponding roll velocity field for each  POD mode.
Note that  the POD modes obtained from the DNS and the RNL simulation 
exhibit  a similar  structure consisting of a streamwise velocity  collocated with a supporting
roll. The variances explained  by the first  three POD modes  are similar  and the structures of the 
modes are also  similar, as shown in Fig. \ref{fig:DNS_RNL_var_270_kx0}, 
although the variance accounted for by the individual modes is not identical.  
The result of importance is the structural similarity of the modes which is indicative of the 
underlying dynamics.

Of dynamical significance is the systematic correlation between the wall-normal velocity $V_s$ of the roll  
and the corresponding streak velocity in these POD modes:  positive $V_s$ is correlated with
low speed streaks (defects in the streamwise  average flow) and vice versa
in all the POD modes.
That all the POD modes exhibit this correlation is consistent 
with the interpretation that the rolls and the streaks form a coherent structure in which the
lift-up mechanism arising from the roll is acting to maintain the streak.  
Consistently, previous work has revealed that the 
Reynolds stress resulting from 
streak-induced organization of turbulence 
in S3T and RNL results in a lift-up process supporting  roll-streaks 
with the same structure as these POD modes
\citep{Farrell-Ioannou-2012,Farrell-etal-2016-VLSM}.   Note that the first 3 DNS POD modes with
$k_x=0$ have roll-streak structure  nearly identical to those in the RNL model
as shown in Fig. \ref{fig:DNS_RNL}.
The wall-normal velocities of the roll component of the  POD modes in both DNS and RNL are about $1/10$ the streak velocity which, assuming
an average non-dimensional   mean flow shear of magnitude 2 (cf.~Fig.~\ref{fig:Mean_flow_Shear})  is consistent with the 
emergence of the associated streak through 
the lift-up mechanism over 5 time units. 

This similarity of the DNS and RNL streamwise-mean POD modes
suggests  that the same dynamics is operating. In the case of RNL this dynamics is known to be that the streaks 
organize the fluctuations so that their associated Reynolds stresses
produce streamwise torque configured to force
rolls collocated with the streak  in such a manner as
to reinforce the preexisting streak by the lift-up process.
This reinforcement 
mechanism is persuasively manifest in the idealized problem of
the instability of 
a background of spanwise homogeneous turbulence to the formation of streamwise streaks.
Statistical state dynamics calculations 
closed at second order  identify this instability, which is the 
fundamental instability underlying the dynamics of turbulence in shear flow, 
by showing that the R-S is the streamwise mean component of  the most unstable eigenfunction
in the SSD.  
Moreover, this unstable eigenfunction has the same form as the POD modes we have identified in both DNS and RNL.
Furthermore, these eigenfunctions   
have the property of destabilizing the roll-streak structure  at all scales, indicating
that the mechanism of streak-roll  formation is intrinsically scale independent \citep{Farrell-Ioannou-2012,Farrell-Ioannou-2017-bifur}.

 
{ 
\section{POD modes of the streamwise-varying
fluctuations from the dominant streak occurring in flow realizations
\label{sec:5}}

\begin{figure*}
\vspace{1cm}
\centering
 \subfloat[]{\includegraphics[width=16pc]{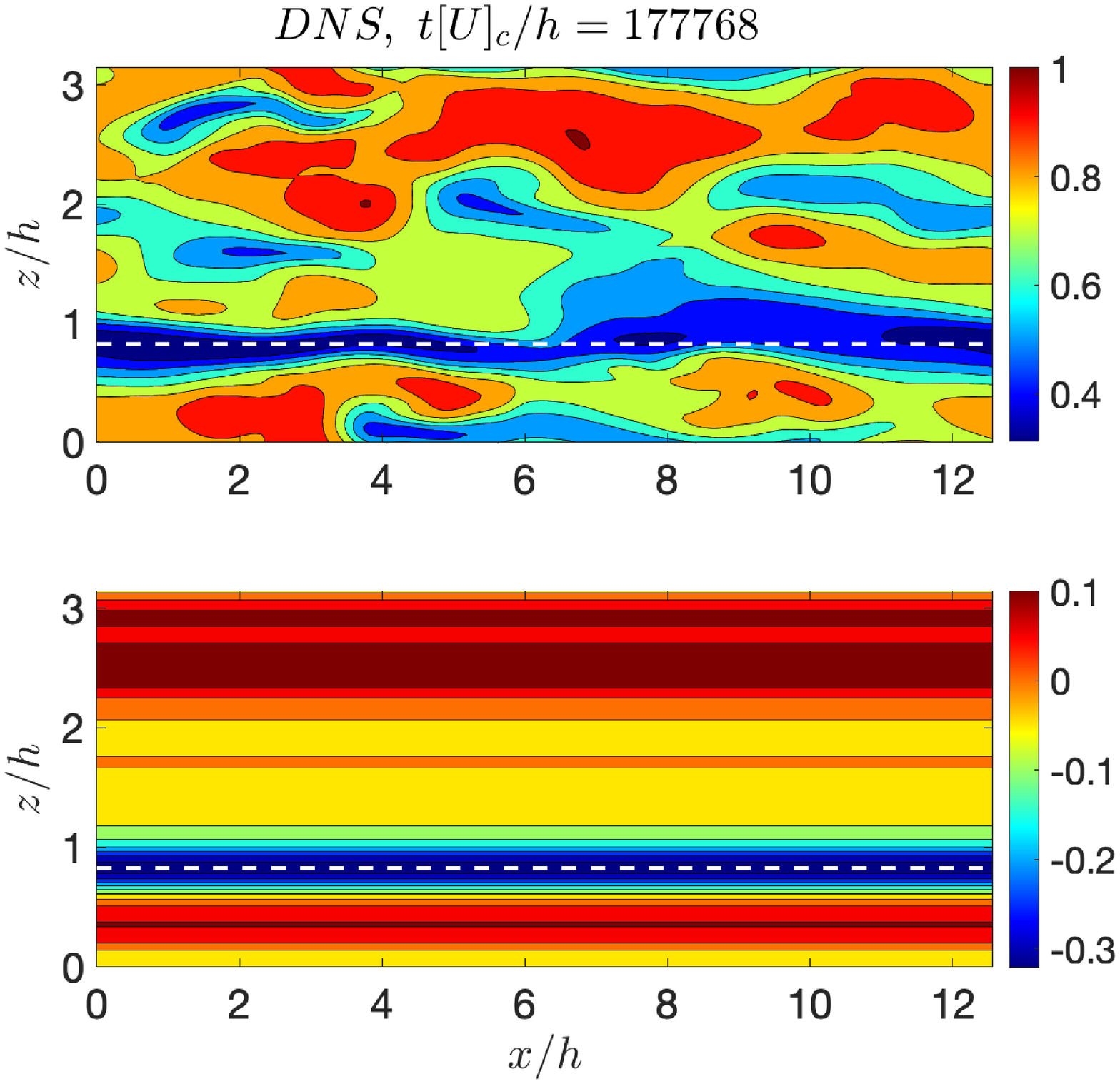}\label{fig:DNS_snap_time}} 
 \subfloat[]{\includegraphics[width=16pc]{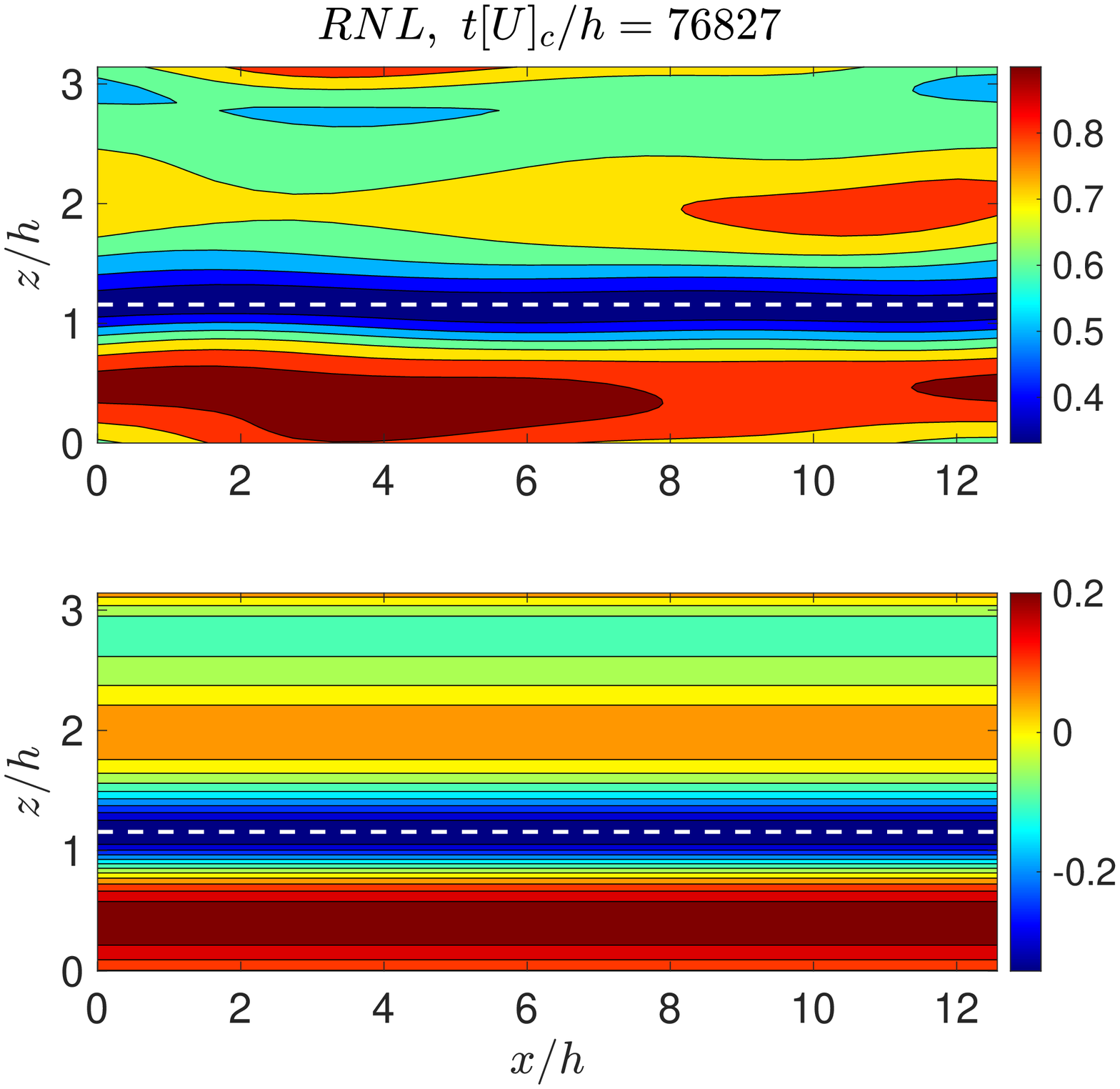}\label{fig:RNL_snap_time}} \\
  \subfloat[]{\includegraphics[width=16pc]{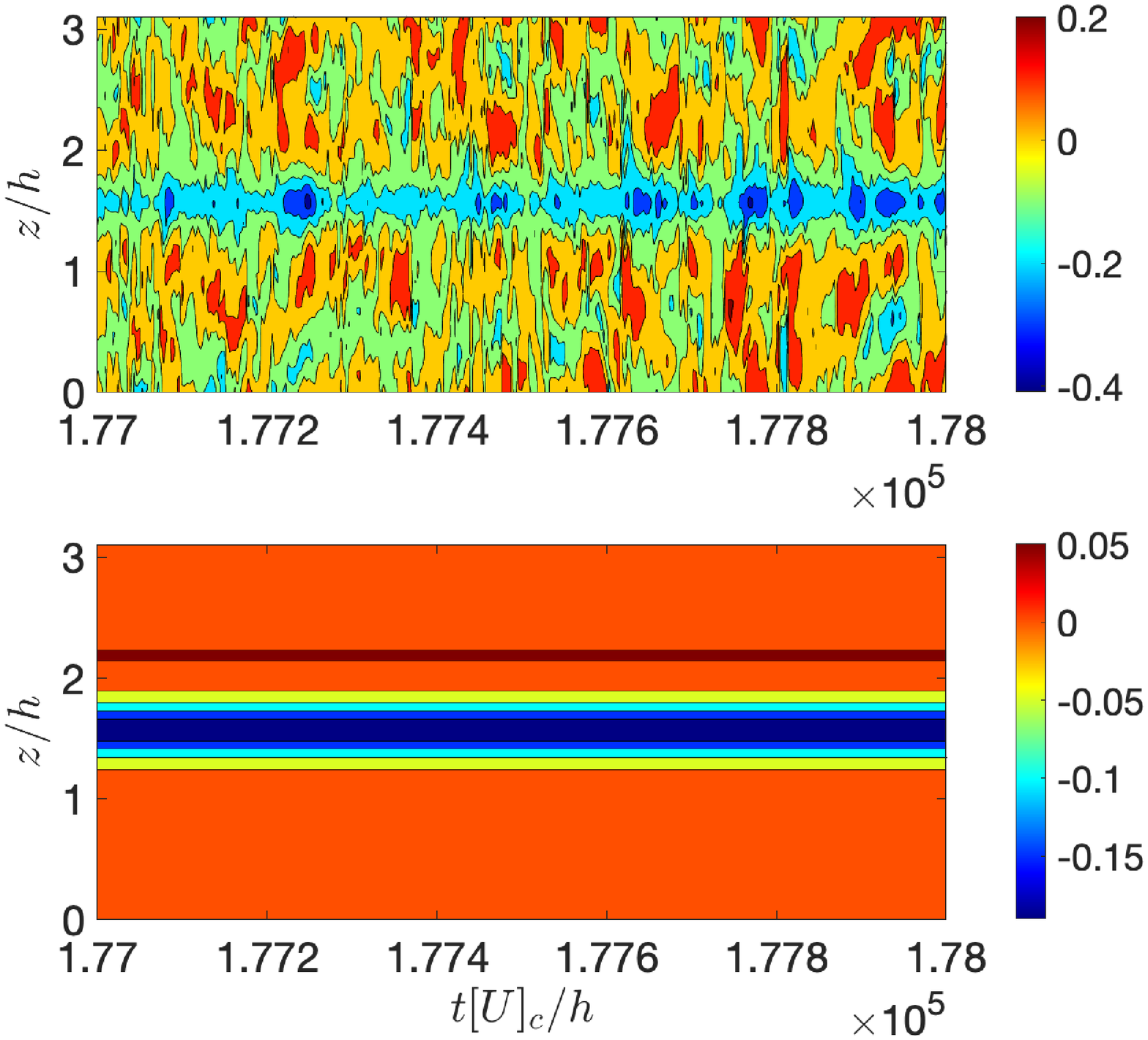}\label{fig:DNS_aligned_time}} 
 \subfloat[]{\includegraphics[width=16pc]{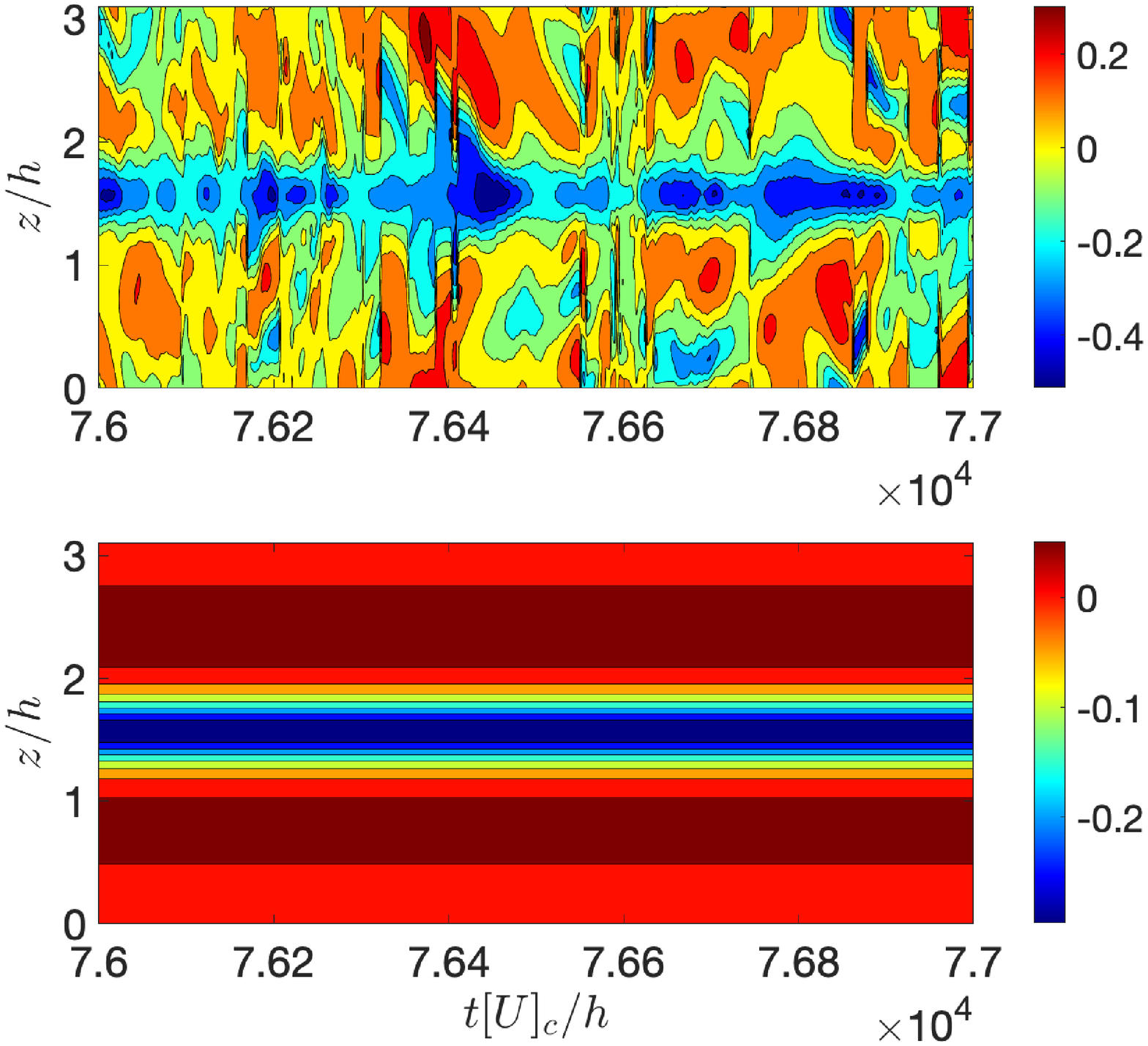}\label{fig:RNL_aligned_time}} 
\caption{(a) Top panel: A snapshot of the streamwise velocity $u$ 
at $t [U]_c/h = 177768$ from  the NL100 simulation  at the wall-normal plane $y/h=0.21$. 
Bottom panel: The $U_s$ component of the above snapshot. The white dashed line in both figures indicates the spanwise location of the $U_s$ minimum. (b) Same as (a) for a snapshot of the RNL100
at $t [U]_c/h = 76827$.
(c) Top panel: A temporal sequence of $U_s$ snapshots for which the streak minima have been
aligned at the channel half-width $z/h=\pi/2$. The total flow snapshot is also subjected to the same shift.
Bottom panel: The ensemble average $U_s$ converges to a negative central 
streak region with weak positive regions on its flanks, whereas the remaining flow is almost spanwise homogeneous.
(d) Same as (c) for the ensemble average $U_s$ of the RNL100 simulation.
} 
\label{fig:ali1}
\end{figure*}


\begin{figure} 
 \vspace{1cm}
  \begin{center}
  \includegraphics[width=1.0\textwidth]{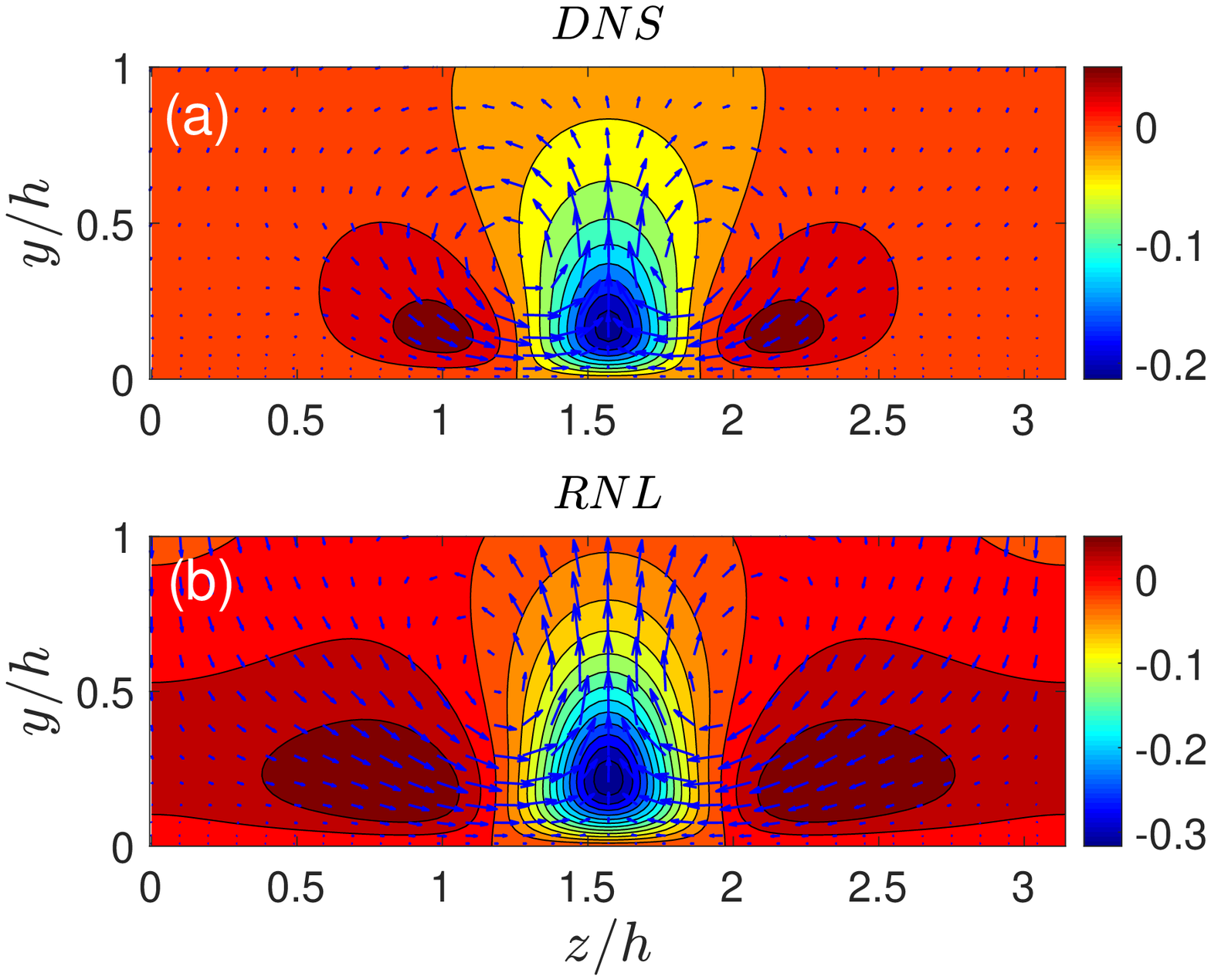}\label{fig:DNS_aligned_avg}  
 \end{center}
 \caption{Contours of the time average collocated  $U_s$ and vectors  of the roll  $(W_s,V_s)$ velocity  on the $(z,y)$ plane 
  for the NS100 (panel(a))  and RNL100  (panel (b)).
  The contour level step is 0.025 in both panels. Panel (a): $\max(|U_s|) = 0.21$, $\max(V_s)=0.024$. Panel (b): $\max(|U_s|) = 0.32$, $\max(V_s)=0.03$.}
 \label{fig:ali2}
\end{figure}



A fundamental dynamical property of turbulence in wall-bounded flows is the spontaneous 
breaking of the spanwise symmetry by the formation of the roll-streak structure.  
Although there is no instability associated with this symmetry breaking in the traditional 
formulation of the NS  using velocity
components for the state, this symmetry is broken by the most 
unstable mode of the simplest nontrivial SSD which is a cumulant expansion closed at second order 
using streamwise-mean velocity and fluctuation covariance for the state variables 
\citep{Farrell-Ioannou-2012,Farrell-Ioannou-2017-bifur}.
While the underlying roll-streak symmetry breaking instability is analytic
in pre-transitional flow analyses made using the S3T SSD, 
the manifestation of this symmetry breaking instability is made imperfect by 
time dependence  both in the pre-transitional and post-transitional DNS and RNL solutions so that the 
roll-streak structure, while prominent,  is randomly spatially displaced 
rather than persisting at  a fixed spanwise location.   Nevertheless, 
the existence of the underlying symmetry break in the spanwise by 
the roll-streak  S3T instability is clearly manifest in the  substantial spatial  extent in 
the steamwise direction and persistence in time of the roll-streak structure
in DNS and RNL, 
indicative of  the underlying analytical  bifurcation.
Informed by the existence of an analytic bifurcation to a time and space  
independent roll-streak structure in pre-transitional flow, we wish to 
isolate structures underlying this fundamental mechanism of roll-streak maintenance 
from the secondary property of 
random variation of the streak location in the spanwise direction.  By this 
simplification we are able to concentrate on the 
interaction of the roll-steak with streamwise fluctuations, which is widely recognized
to be associated with the maintenance
of turbulence, although the dynamics of this interaction remains controversial 
(cf. \cite{Jimenez-Moin-1991,Hamilton-etal-1995,Waleffe-1997,Jimenez-Pinelli-1999,
Schoppa-Hussain-2002,Farrell-Ioannou-2012, Farrell-Ioannou-2017-sync,Farrell-etal-2016-PTRSA,Lozano-Duran-etal-2021}).
A point of agreement of the above studies 
is that the streak and fluctuations are collocated to form a  dynamical structure. 
Thus an  accurate statistical description  of the $k_x\ne0$ structures  
will be sought 
by  performing at every instant of  time a  spanwise translation of the entire flow field data so that the dominant streak together with  its associated
fluctuations is  at the center of the channel, $z/h=\pi/2$.

A reliable indicator of the streak location is the spanwise  $z/h$ 
coordinate of the $\min(U_s)$ associated with the dominant low speed streak structure (Figs. \ref{fig:DNS_snap_time} and \ref{fig:RNL_snap_time}). 
We proceed to identify this location in the
flow realizations by finding the $z$ coordinate of this minimum velocity at a fixed distance from the wall, $y/h=0.21$, 
where the $|\min(U_s)|$ attains  it's largest values, 
and translate the total flow field  so  that the $U_s$ minima occur  at the  same spanwise location at the center of the channel at $z/h=\pi/2$,
 while retaining the time order.
The effect of this operation on the streamwise average $U_s$ velocity is shown in the top panel of Figs. \ref{fig:DNS_aligned_time} and \ref{fig:RNL_aligned_time} for NL100 and RNL100 respectively. The modified time-series of the $U_s$ produce an aligned slow speed region at $z/h=\pi/2$ in both cases, while further away from this core region the uncorrelated
high and low speed streaks cancel out.  
The streamwise-mean streak, $U_s$, on the plane $y/h=0.21$ 
resulting from this procedure  is shown in
the bottom panel of Figs. \ref{fig:DNS_aligned_time} and \ref{fig:RNL_aligned_time}.
The  structure in the $y-z$ plane  of the roll-streak for NL100 and RNL100    is shown in Figs. \ref{fig:ali2}a,b
using contours for $U_s$ and  vectors for  $(W_s,V_s)$.

\begin{figure} 
 \vspace{1cm}
  \begin{center}
    \includegraphics[width=0.5\textwidth]{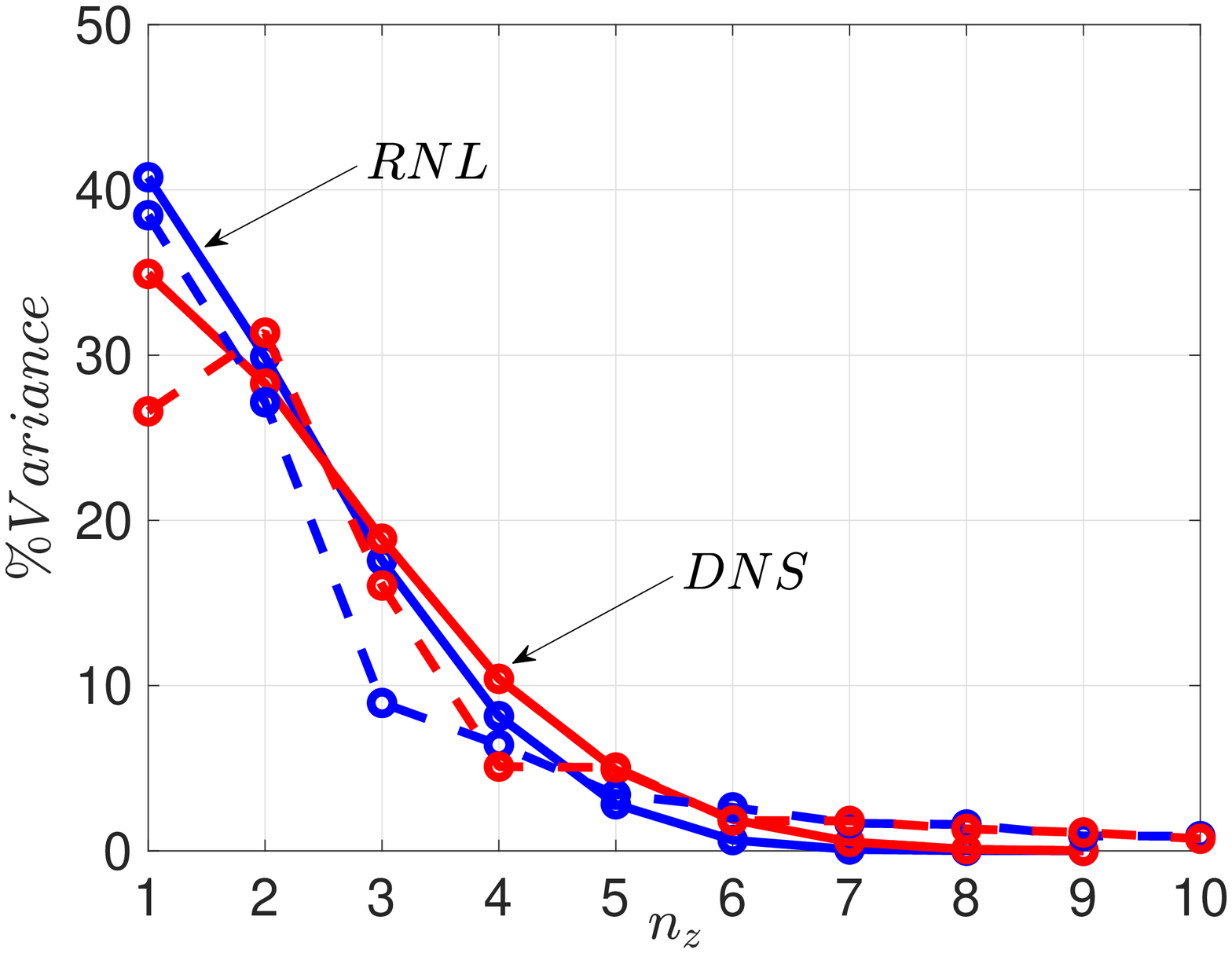}
 \end{center}
 \caption{The percentage variance accounted for by the first  Fourier spanwise components of the mean streaks in Fig. \ref{fig:ali2}.
 Solid red line: for the mean streak of the DNS, solid blue line: for the mean streak of the RNL.
 The dashed lines with the corresponding colors show the percentage  variance of the corresponding POD modes  with spanwise wavenumbers $n_z=1,..,10$.} 
 \label{fig:FC}
\end{figure}

 \subsection{Relating POD modes to roll-streak  structures in the flow}

\begin{figure} 
 \vspace{1cm}
  \begin{center}
    \includegraphics[width=1\textwidth]{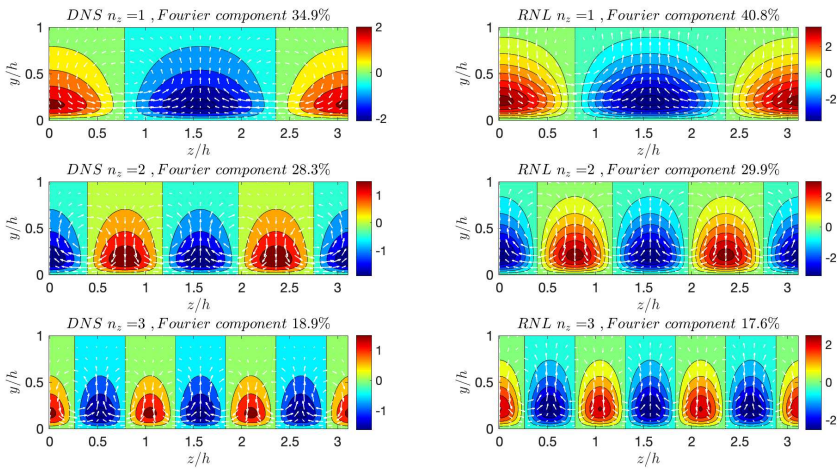} 
 \end{center}
 \caption{Contours of the  $U_s$ and vectors  of the roll  $(W_s,V_s)$ velocity  on the $(z,y)$ plane of the 
 first three  spanwise  Fourier components of the mean streaks of Fig. \ref{fig:ali2}. Left column of  the DNS, right column of the RNL.
  The contour level step is 0.2 in all panels.  } 
 \label{fig:FC1}
\end{figure}

There are alternative explanations for the striking appearance of POD 
modes for $k_x=0$ which differ in spanwise wavenumber while exhibiting roll-streak structure (cf. Fig. \ref{fig:DNS_RNL}).  
One interpretation of these structures is that they correspond to stable
linear S3T modes that are  maintained by fluctuations in the homogeneous background of turbulence.  
Due to the scale insensitivity of the roll-streak formation process,  a spectral 
hierarchy of self-similar roll-streak structures are supported as 
modes by the turbulence as revealed by S3T \citep{Farrell-Ioannou-2012}.  
These roll-streak modes have different scales and damping 
rates and are therefore expected to be excited at different amplitudes.
 This interpretation of the POD modes is appropriate in the case 
of the roll-streaks that emerge in pre-transitional flows  as discussed in \cite{Farrell-Ioannou-2017-bifur}.
Also, in beta-plane turbulence one observes intermittent emergence of jets 
with structure corresponding to stochastically excited 
S3T modes, manifestations of which are referred to in observations as latent jets
\citep{Constantinou-etal-2014,Farrell-Ioannou-2019-book}.
In this interpretation of the POD modes with various spanwise wavenumber, the POD modes are 
identifying structures that are  regarded in the traditional manner as being independent 
harmonic structures in $z$, as is appropriate for a homogeneous coordinate in the flow.

However,  there is an alternative interpretation, which is that the dominant structure in the flow is
 the finite amplitude localized low-speed streak of Fig. \ref{fig:ali2}.  In this interpretation  
the POD modes reflect the amplitude and structure of the spanwise Fourier components collocated to comprise the Fourier synthesis 
of this R-S structure
rather than corresponding to structures with independent existence. The assumption underlying 
this interpretation is that a mode has arisen in the flow that results in a spontaneous 
symmetry breaking; in the example under analysis here the S3T R-S instability is implicated.

In order to determine which of these alternative explanations for the POD structure at $k_x=0$ is correct
we obtain the spanwise Fourier decomposition of the three velocity components of the mean streak and roll velocities shown in  Fig. \ref{fig:ali2}.
We now compare the spanwise Fourier decomposition of the R-S 
(cf. Fig. \ref{fig:ali2}) with the POD modes which have necessarily Fourier structure in $z$.  
The percentage variance accounted for by the first 10 spanwise harmonics in the Fourier decomposition 
of the streak and the percentage variance of the POD modes with the same 
spanwise wavenumber are compared in Fig. \ref{fig:FC}.  
These spectra are similar enough to suggest that the POD modes are  the Fourier components of 
the streak and indeed, 
if the POD modes are  used to compose a structure by using the corresponding POD 
variances collocated at zero phase, this produces a close 
correspondence to the streak in Fig.  \ref{fig:ali2} (not shown).  Consistent with this 
interpretation is the comparison of structure between the POD modes shown in Fig. \ref{fig:DNS_RNL}
and the Fourier modes of the streak shown in Fig. \ref{fig:FC1}.
This agreement is expected given that in our DNS $65\%$ of the $k_x=0$ fluctuation energy is accounted for by the mean streak structures
in Fig. \ref{fig:ali2}  (the low speed streak accounts for $40 \%$  and the high-speed streak $25 \%$ (not shown), the RNL percentages are similar).
The agreement shown in these figures leads us to conclude that the second of these explanations, that  the POD 
spectrum arises primarily from Fourier decomposition of the low-speed R-S, is correct.
In summary we conclude that while  POD analysis is consistent with identification of independent R-S 
structures, the alternative interpretation that the 
POD modes rather identify the individual components comprising the 
Fourier synthesis of a nonlinearly equilibrated localized coherent 
structure with complex R-S form, requiring many Fourier modes in its representation, is clearly preferable.

\subsection{Determining the $k_x \ne 0$ POD modes associated with the collocated low-speed streak flows}

\begin{figure} 
 \vspace{1cm}
  \begin{center}
    \subfloat[]{\includegraphics[width=0.5\textwidth]{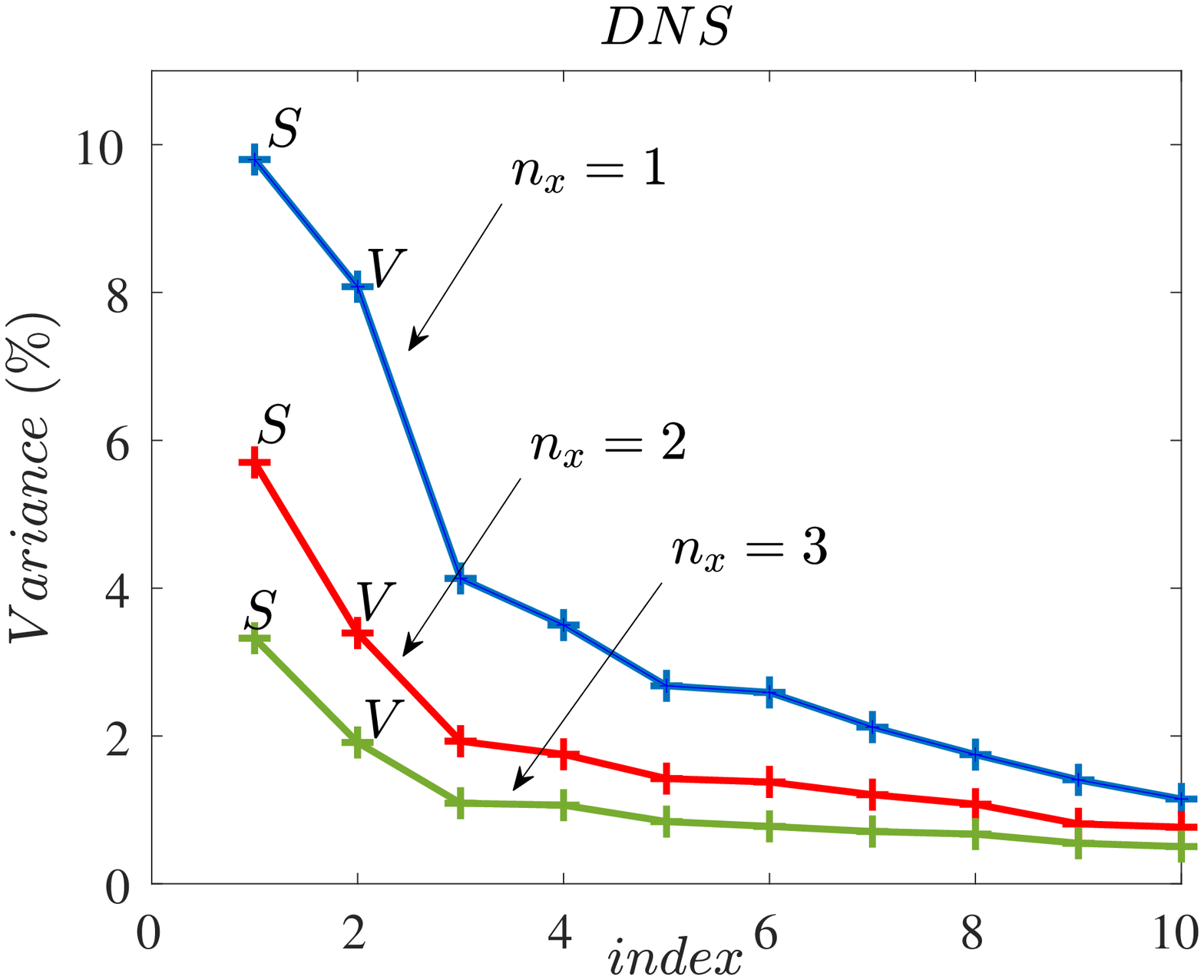}\label{fig:dns_var_tuk}} 
    \subfloat[]{\includegraphics[width=0.5\textwidth]{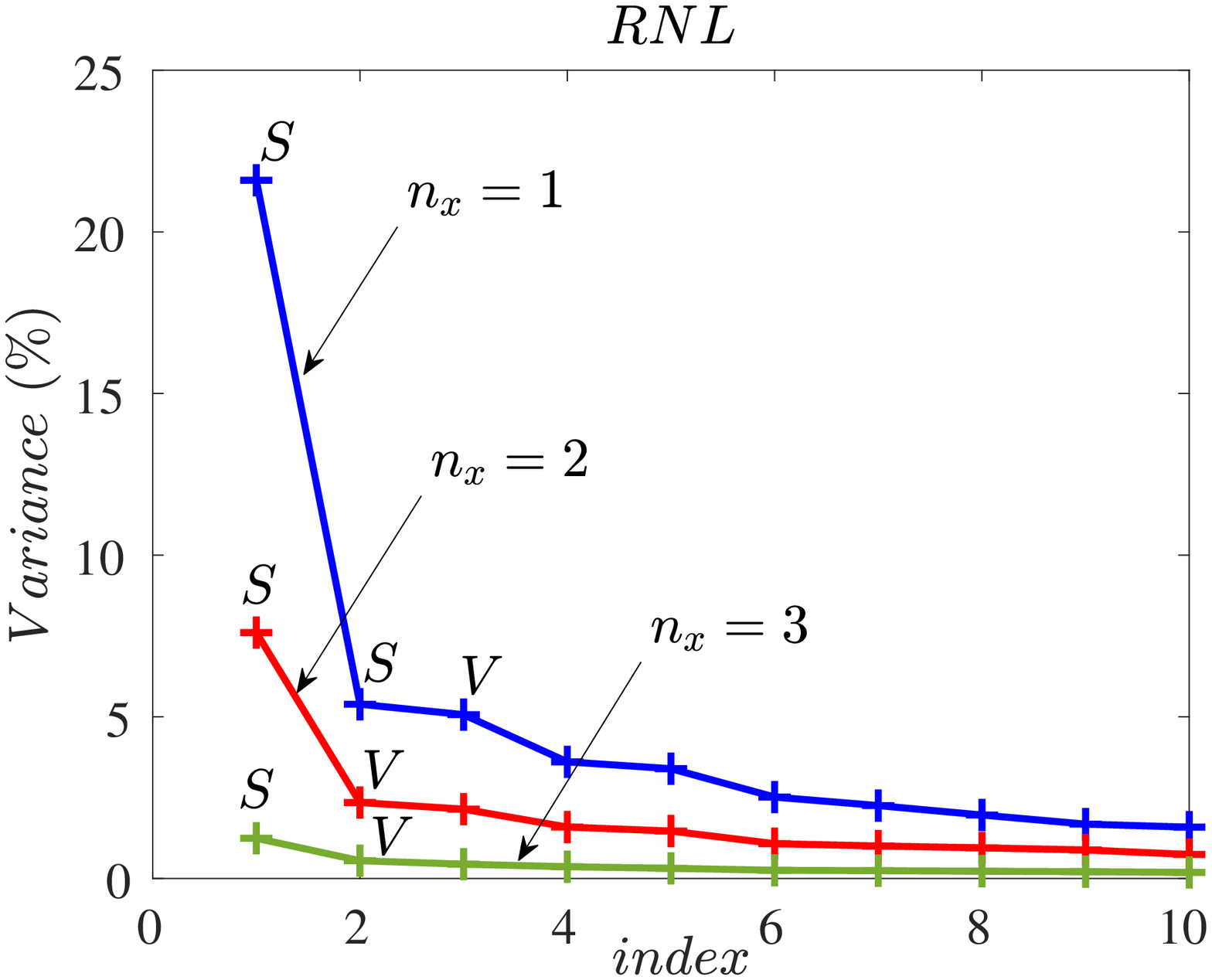} \label{fig:rnl_var_tuk}}
 \end{center}
 \caption{Percentage variance  accounted for by the POD modes as a function of the order of the mode; (a) in NL100 and 
 (b) in RNL100.  POD modes  with streamwise Fourier component $n_x=1$  are in blue; those with streamwise Fourier component $n_x=2$ are in red; and those with streamwise Fourier component $n_x=3$ in green. The sinuous modes are indicated with an S, the varicose with a V. The corresponding streamwise wavenumber is $k_x=2 \pi n_x/L_x$.}
\label{fig:pert_var}
\end{figure}


Having isolated the streamwise mean R-S structure 
in DNS and RNL and identified the $k_x=0$
 POD modes as consistent with Fourier synthesis of this coherent structure,
we turn now to exploiting 
POD analysis to obtain and interpret dynamically the
fluctuations about the mean flow containing  the streak structures of Fig. \ref{fig:ali2}. 
We  first
Fourier decompose the fluctuation velocity ${\cal{U}}'= [ u(\xv,t), v(\xv,t), w(\xv,t)    ]^T$ in $x$
so that ${\cal{U}}'= {\cal{U}}'_{k_x}(y,z) e^{i k_x x}$ 
The POD modes are obtained from eigenanalysis of the two point spatial covariance:
\be 
C_{k_x}(y_i,z_i,y_j,z_j) = \< {\cal U}'_{k_x}(y_i,z_i) {\cal U}'^{\dagger}_{k_x}(y_j z_j) \>,
\ee
where 
$\langle \cdot \rangle$ denotes the time mean, and $\dagger$ indicates the Hermitian transpose. 
Each POD mode is of the form $ 
\left[ \begin{array}{c}
  \alpha_{k_x}{(y,z)}, 
   \beta_{k_x}{(y,z)},
   \gamma_{k_x}{(y,z)}
\end{array}
\right]^T e^{i k_x x}
$, with
$\alpha_{k_x}(y,z)$, $\beta_{k_x}(y,z)$ and $\gamma_{k_x}(y,z)$ determining the $(y,z)$ spatial
structure  of  the  velocity components of the POD mode. 
Note that the $n_x$ component of the velocity field has streamwise wavenumber $k_x= n_x \alpha$,



The  flow fields shown in Fig.  \ref{fig:ali2} 
reveal  spanwise localized  R-S structures 
symmetric about $z/h=\pi/2$. 
In order to isolate  the streamwise-varying  POD modes associated with the localized low-speed streaks
while  avoiding contamination by the far-field 
we weight the data used to calculate  the covariances $C_{k_x}$  with 
a spatial filter that suppresses  the variance in the far-field.
We have chosen  a  Tukey  filter in the interval $z=[0,\pi]$ with equation:
\begin{equation}
f(z) =\begin{cases}
{1}/2 \left [ 1+\cos \left (\pi/\delta \left ((\pi-2 z)/\pi +  (1 -\delta) \right )\right )\right ]~,& (\pi-2 z)/{\pi}<\delta-1,\\
1~,&   |(\pi-2 z)/{\pi}|\le 1-\delta~,\\
{1}/{2} \left [1+\cos \left (\pi/\delta \left ((({\pi-2 z})/{\pi}-(1 -\delta)  \right) \right )\right ]~,& (\pi-2 z)/{\pi}> 1-\delta ~.
\end{cases}
\end{equation}
The parameter $\delta$ dictates the width of the filter and is chosen to sample the fluctuation field
in the vicinity of the streak. The values $\delta=0.7$ and $\delta=0.55$ are selected for NL100 and RNL100, respectively, since the 
RNL100 streak covers a wider area of the spanwise flow.  


 \subsection{Results  for the POD modes with $k_x  \ne 0$ associated with the collocated low-speed streak flows}
 
 The energy density 
accounted for by the  first 10 POD modes  for each of  the first three $n_x$ wavenumbers
is shown  in Fig. \ref{fig:dns_var_tuk} for NL100 and in 
Fig. \ref{fig:rnl_var_tuk} for  RNL100.
Similarly, in Fig. \ref{fig:dns_k1s},  Fig. \ref{fig:dns_k2s} and  Fig. \ref{fig:dns_k3s} are shown the corresponding structures
of the first three sinuous POD modes
with streamwise wavenumbers $n_x=1,2,3$.
Note that the dominant POD modes in both DNS and RNL 
are characterized by  a similar intricate complex three dimensional structure
that reflects the complexity of the underlying dynamics.

In the $y-z$ cross sections  (top panels of Fig. \ref{fig:dns_k1s},\ref{fig:dns_k2s},\ref{fig:dns_k3s}) the POD modes exhibit streamwise streaks produced by lift-up which 
are seen to be coincident with supporting roll circulations. 
Similar 
roll circulation and associated streak structures 
were educed  from analysis of DNS data to arise in association with  sweep and ejection events 
by  \cite{Lozano-Duran-etal-2012} 
(cf. their Fig. 12d) and \cite{Encinar-Jimenez-2020}  (cf. their Fig. 12).
Note that these streaks and associated  rolls are located at the flanks of the central streamwise  streak
and are harmonic in the streamwise direction and should not be confused with the 
entirely different  roll-streak  structure  that  arise from lift-up induced by the Reynolds stresses of the fluctuations
and, which in contrast, force the central streak with $k_x=0$.

In the $x-y$ cross sections (middle panels of Fig. \ref{fig:dns_k1s},\ref{fig:dns_k2s},\ref{fig:dns_k3s}) the POD modes exhibit the tilted structure indicative of  linear amplification  by the Orr
mechanism. This characteristic Orr structure has been associated   with  sweep and ejection events
in wall-turbulence by
\cite{Encinar-Jimenez-2020}  (cf. their Fig. 1) and evolution of these structures was found  to accord with the linear evolution
of  optimal perturbations with  Orr form on the cross-stream shear \citep{Jimenez-2013lin,Encinar-Jimenez-2020}.

In the $x-z$ cross sections (bottom panels of Fig. \ref{fig:dns_k1s},\ref{fig:dns_k2s},\ref{fig:dns_k3s}) the POD modes exhibit orientation with the streak indicative of an energy extracting
sinuous  oblique wave collocated with the 
streak. Sinuous structure  of streak fluctuations
has been associated with inflectional instability
 \citep{Waleffe-1997}  and  with  optimally growing 
perturbations \citep{Schoppa-Hussain-2002}.

%
 
  The dominance of the top sinuous structure variance, shown in Fig \ref{fig:dns_var_tuk}, indicates that it is 
 preferentially expressed
 relative to the other components of the fluctuation field in both DNS and RNL. 
 The second in variance POD mode
 is usually varicose and it also exhibits a similar structure in DNS and RNL (not shown). The sinuous and 
 varicose ordering  of the POD modes
 is indicated in Figures  \ref{fig:dns_var_tuk} and  \ref{fig:rnl_var_tuk}.


Qualitative agreement  in structure between the top   POD modes of the DNS 
and RNL is apparent. 
Despite differences such as the greater extent in $y$ of the streak structure of the RNL POD modes 
(cf. top panels of Figures 
\ref{fig:dns_k1s}, \ref{fig:dns_k2s} and \ref{fig:dns_k3s}),  the compelling similarity  in  structural features,
 and the phasing  between fluctuation fields revealed by  the DNS and RNL  POD modes
argues for the operation of  a parallel dynamics in these two systems.
The exact statistical steady state, including the amplitude and spatial extent of the similar structures, is 
determined by the  non-linear  feedback regulation
between the mean and fluctuation fields  \citep{Farrell-Ioannou-2012}.
In particular, the  feedback regulation produces an   RNL equilibrium with a 50\% greater spatial  extent
in the fluctuation streak shown in the upper panels, which is consistent with a similarly greater $V$ velocity in the RNL (cf. Fig. \ref{fig:ali2}). 
While similarity of the R-S structure between DNS and RNL is manifest, there is no simple argument 
for the  exact amplitude of the  mean velocity components that the associated non-linear regulator  settles on.
Intuitively this can be understood from considering that the feedback regulation 
is using the Reynolds stresses to adjust the R-S to Lyapunov 
neutrality and this can be accomplished by adjusting the amplitude, structure and temporal 
variation of the R-S in many ways so that a single easily predicted equilibrium structure is not to be expected.

\begin{figure} 
 \vspace{1cm}
  \begin{center}
    \includegraphics[width=1\textwidth]{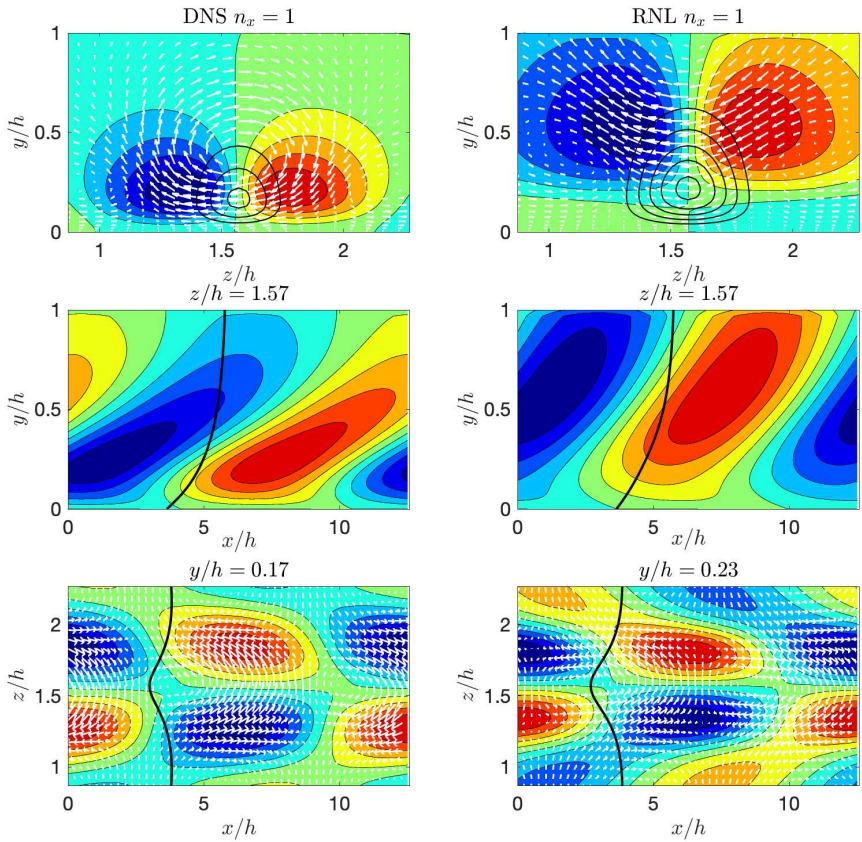} 
 \end{center}
 \caption{The first sinuous POD mode with streamwise Fourier component $n_x=1$ in  NL100 (panels on the left) and in RNL100 (panels on the right). 
 Top panels: contours of the $u$ velocity of the POD mode in the $z-y$ plane at $x=0$ and 
 vectors of  $(w,v)$ velocity on this plane.
 Middle panels: contours of the $w$ velocity  in the $x-y$ plane at the center of streak where the $u$ and $v$ velocities vanish.
 Bottom panels: contours of the $v$ velocity in the $x-z$ plane at the center of streak and 
 vectors of  $(u,w)$ velocity on this plane.  The mean flow structure is indicated by the solid back line. The black contours
 in the top panels  show the streak  contours in the interval $[-0.35,-0.1]$ at contour intervals of $0.05$. All other quantities have been normalized to 1 and the contour levels  is $0.2$.
 The first sinuous DNS POD mode
 accounts for $9.8\%$ of the total variance of the streamwise varying velocity fluctuations of the flow (which includes all $k_x\ne0$), while the first 
 sinuous RNL POD mode accounts for  $21.6 \%$  of the total fluctuation variance (cf. Fig. \ref{fig:pert_var}). }  
 \label{fig:dns_k1s}
\end{figure}

\begin{figure} 
 \vspace{1cm}
  \begin{center}
 \includegraphics[width=1\textwidth]{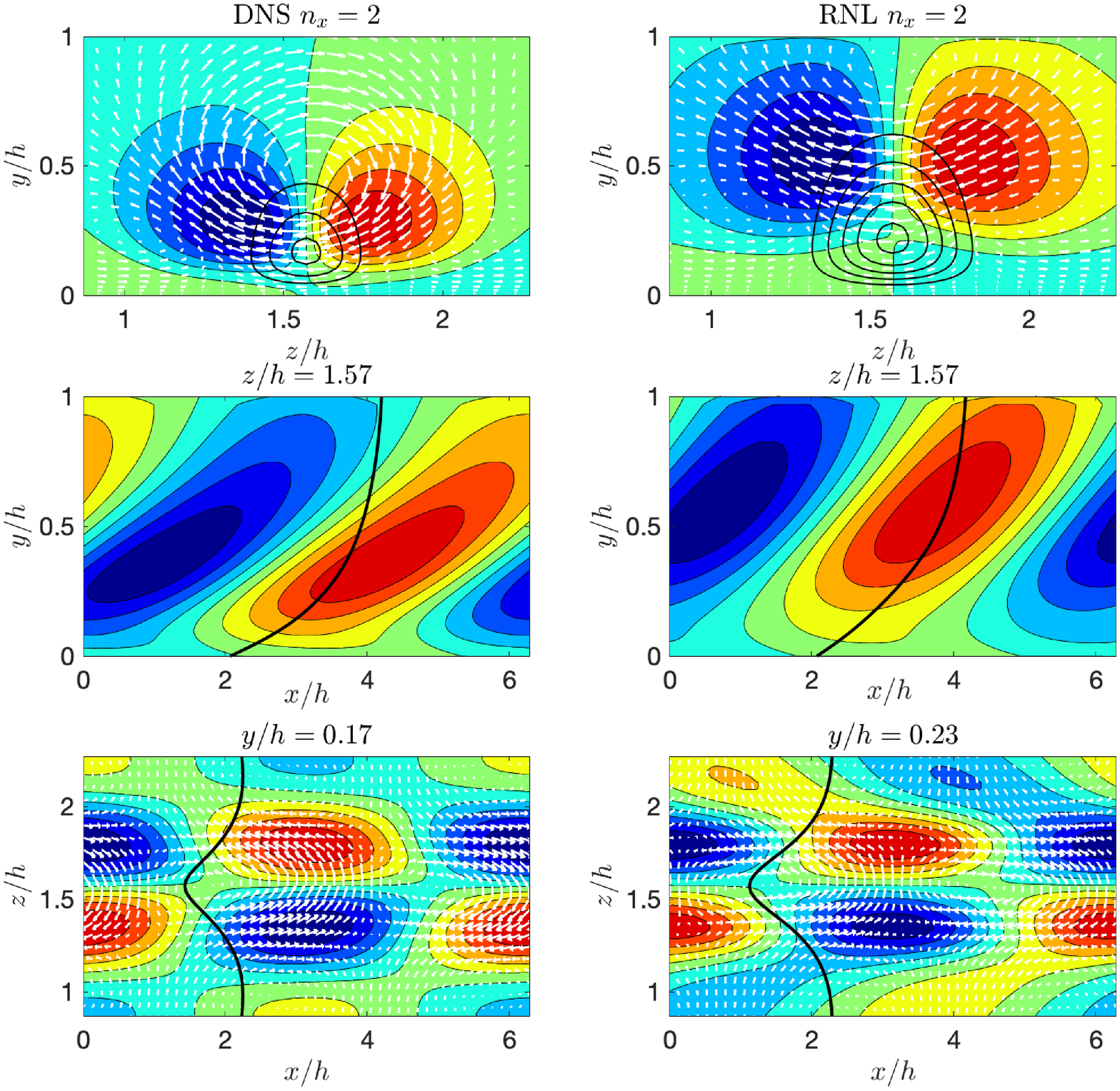}
 \end{center}
 \caption{As in Fig. \ref{fig:dns_k1s} for the  first sinuous POD mode with streamwise Fourier component $n_x=2$. A single streamwise wavelength of the POD mode has been plotted. The first sinuous DNS POD mode (which is the first in variance POD)
  accounts for $5.7\%$  of the total variance of the streamwise varying velocity fluctuations of the flow, while the first 
 sinuous RNL POD mode (which is also the first  in variance POD) accounts for  $7.6 \%$  of the total fluctuation variance.}  
 \label{fig:dns_k2s}
\end{figure}


\begin{figure} 
 \vspace{1cm}
  \begin{center}
 \includegraphics[width=1\textwidth]{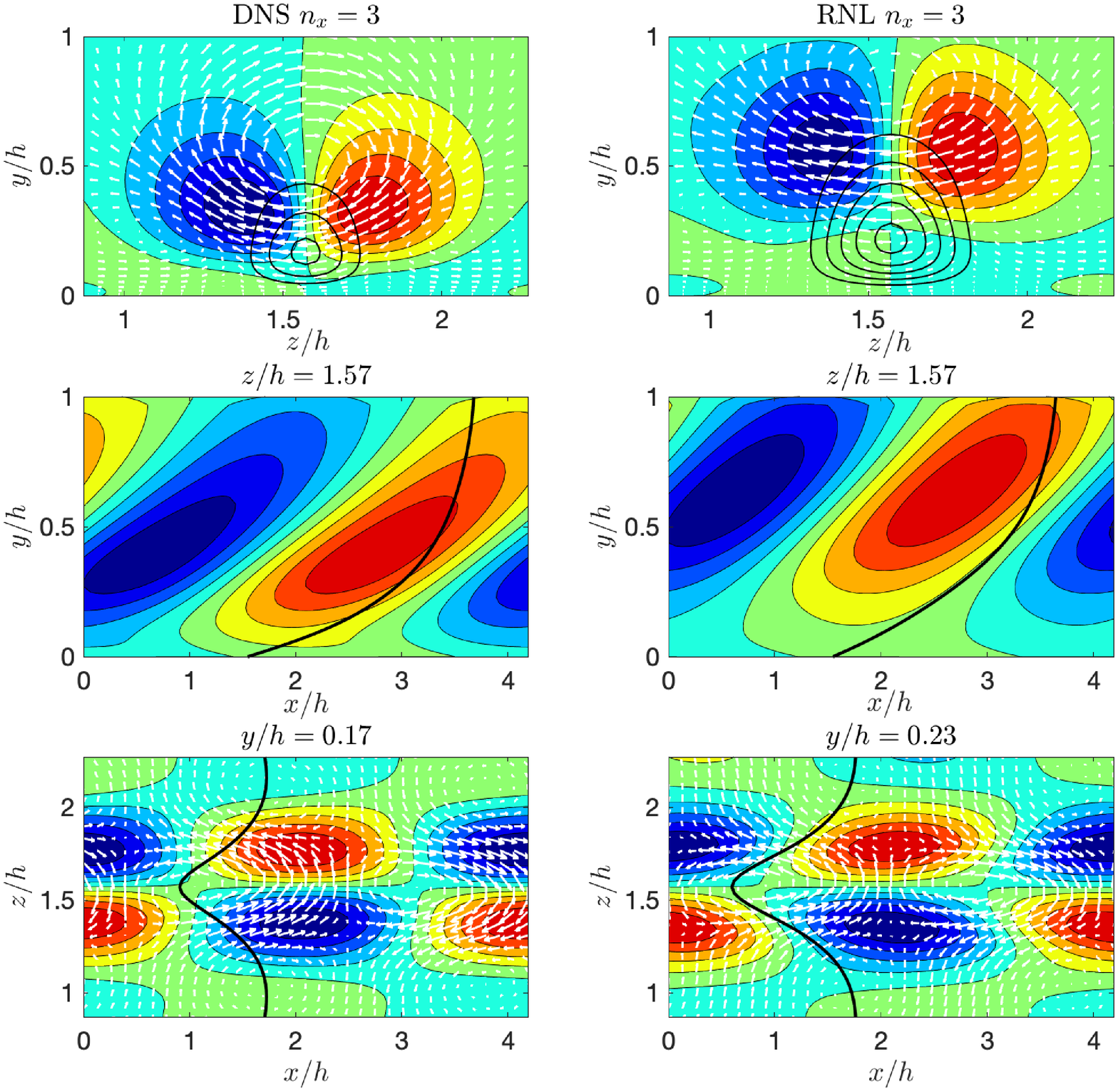}
 \end{center}
 \caption{As in Fig. \ref{fig:dns_k1s} for the  first sinuous POD mode with streamwise Fourier component $n_x=3$. A single streamwise wavelength of the POD mode has been plotted. 
  The first sinuous DNS POD mode (which is the first in variance POD)
  accounts for $5.7\%$  of the total variance of the streamwise varying velocity fluctuations of the flow, while the first 
 sinuous RNL POD mode (which is also the first  in variance POD) accounts for  $1.2 \%$  of the total fluctuation variance.}  
 \label{fig:dns_k3s}
\end{figure}

\begin{figure} 
 \vspace{1cm}
  \begin{center}
    \includegraphics[width=1\textwidth]{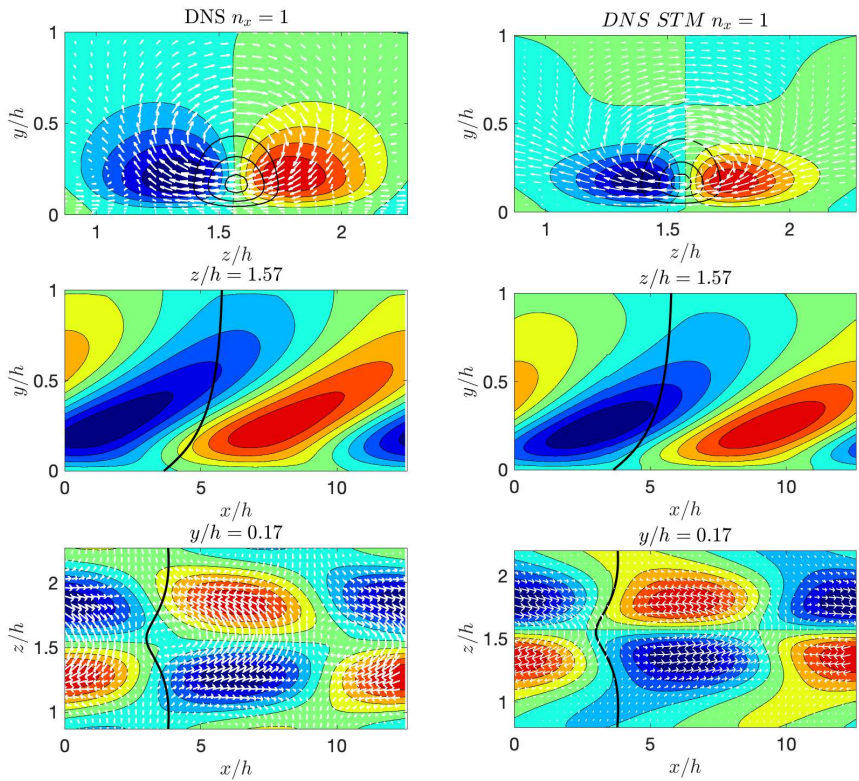} 
 \end{center}
 \caption{
 Comparison of the first sinuous  POD mode in  NL100  with streamwise Fourier component $n_x=1$ (panels on the left)  with the first
 sinuous  and first  POD mode of the STM  with $T_d=30$ on the  DNS mean low-speed streak shown in Fig.
 \ref{fig:ali2}a  (panels on the right). 
 The velocity fields are as in Fig. \ref{fig:dns_k1s}.
   The POD with the largest variance is the sinuous mode 
  in both DNS and STM.}  
 \label{fig:dns_stm_1}
\end{figure}

\begin{figure} 
 \vspace{1cm}
  \begin{center}
    \includegraphics[width=1\textwidth]{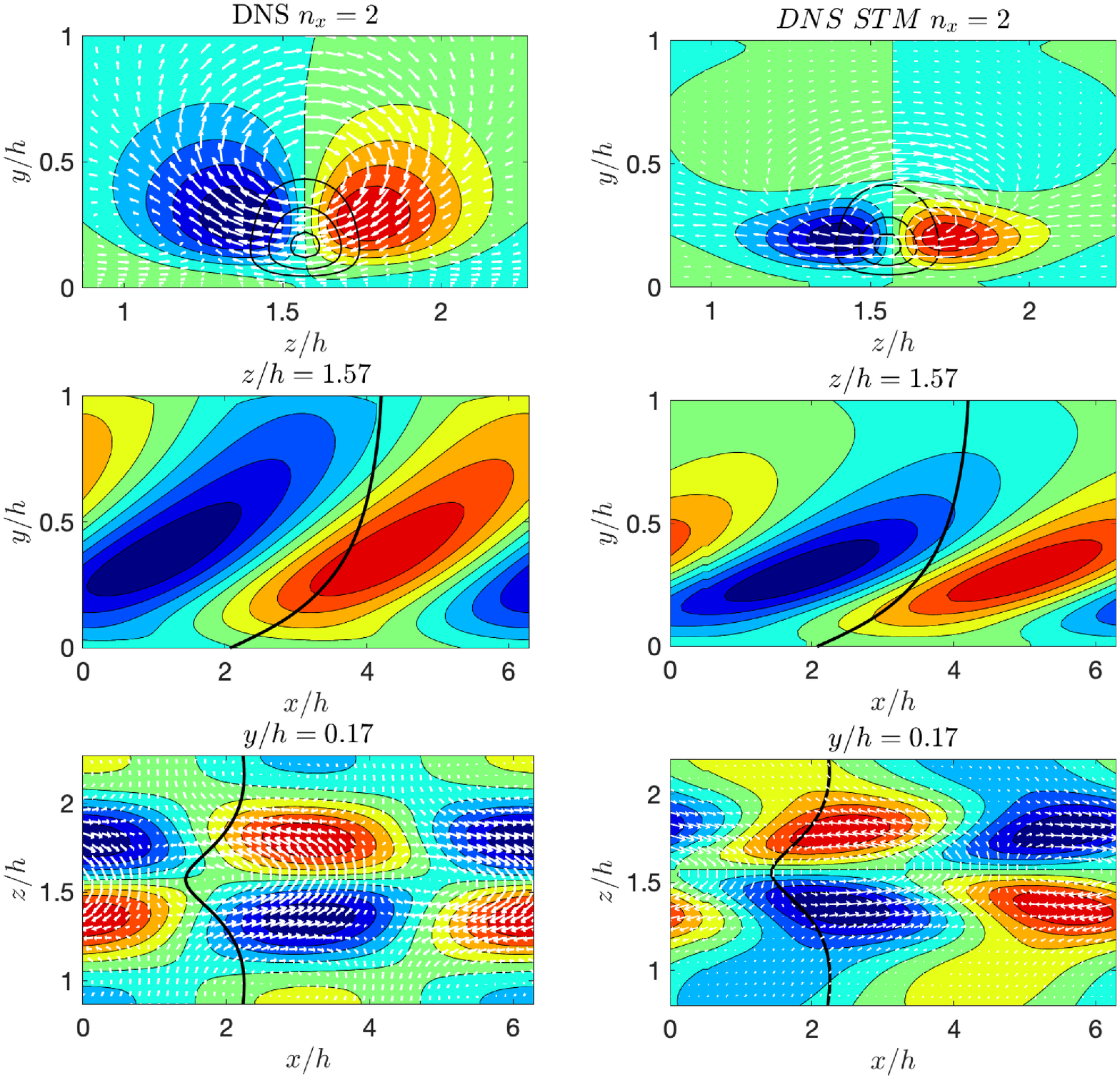} 
 \end{center}
 \caption{
 As in Fig. \ref{fig:dns_stm_1} for $n_x=2$ fluctuations. A single streamwise wavelength of the POD mode has been plotted.}  
 \label{fig:dns_stm_2}
\end{figure}

\begin{figure} 
 \vspace{1cm}
 \begin{center}
    \includegraphics[width=1\textwidth]{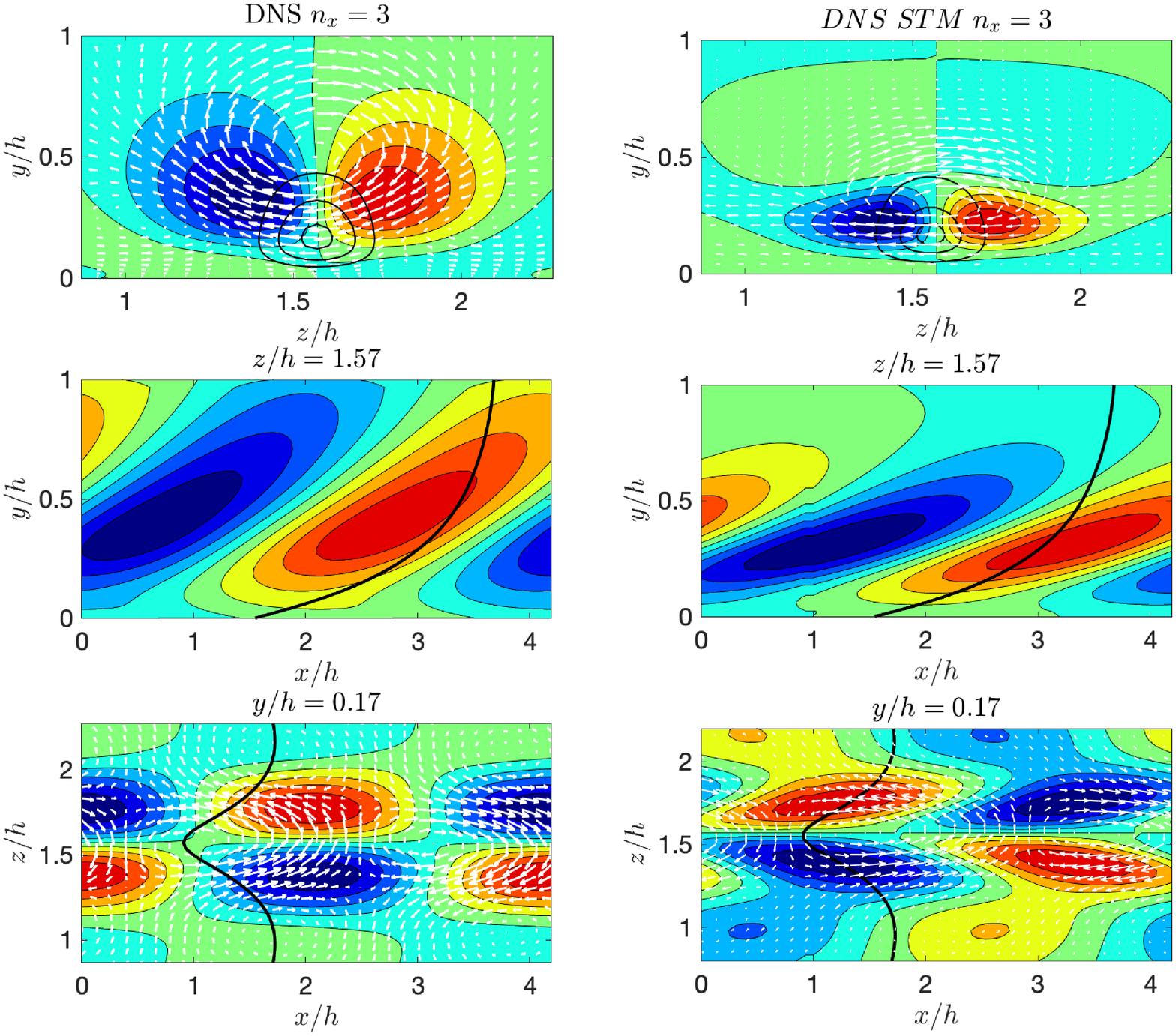} 
 \end{center}
 \caption{
 As in Fig. \ref{fig:dns_stm_1} for $n_x=3$ fluctuations.  A single streamwise wavelength of the POD mode has been plotted.}  
 \label{fig:dns_stm_3}
\end{figure}

\subsection{Relation  of  the  POD modes of the fluctuations in DNS  to the POD modes  
of the fluctuations in a linear  stochastic turbulence model}

As remarked earlier, the striking structural similarity in the  POD modes of fluctuations on
the streamwise-mean streak  in 
DNS and RNL suggests a common dynamical origin.  Given that the streak is modally stable, 
the default explanation would be excitation of 
transiently growing  fluctuations to the streamwise streak by 
the  turbulent background velocity field. 
Transient growth of optimal perturbations on a cross-stream shear has
recently been shown to reasonably accord with the short time evolution
of the structures that arise during sweep-ejection events \citep{Encinar-Jimenez-2020}.
These structures, when  evolved,  assume the characteristic tilt of the POD modes shown  in
the middle panels of Figures \ref{fig:dns_k1s}, \ref{fig:dns_k2s} and \ref{fig:dns_k3s}.
Within the context of POD analysis the appropriate extension of the optimal transient growth analysis of
 \cite{Encinar-Jimenez-2020} would be 
calculation of the POD modes arising from  the ensemble mean fluctuation covariance 
excited by the background turbulence.

%


This can be implemented by calculating the covariance  using the stochastic turbulence model (STM) 
governed by the equations:
 \begin{equation} 
\partial_t\u +   U \partial_x \u +
( \u \bcdot \nabla ) U~\widehat{\xv}  + \nabla p-  R^{-1} \Delta  \u
= {\bf f}~,~~\nabla \bcdot \mathbf{u} = 0~,
 \label{eq:STM}
 \end{equation}
 with no slip boundary conditions at the channel walls and periodic boundary conditions in $x$ and $z$,
 where $U(y,z) \widehat{\xv}$ is  the equilibrium low-speed streak of the DNS (or the RNL).
The equations are the fluctuation equations of the DNS \eqref{eq:NSp} in which the fluctuation-fluctuation
nonlinearity has been  replaced by a state independent forcing, ${\bf f}$, white in space and time. 
This  simplest parameterization imposes the  least 
a priori assumption on the structures arising from the dynamics and
is adequate for our purpose. The time dependent 
streamwise-mean flow, $\U$, of the DNS has been replaced by the time-independent flow, $U(y,z) \widehat{\xv}$,
obtained by collocating and averaging the low-speed streak. This  mean streak is shown in Fig. \ref{fig:ali2}.
%

 Stability of the mean flows shown in Fig. \ref{fig:ali2} assures  the asymptotic approach of the covariance
 of the fluctuations governed by \eqref{eq:STM}  to a  statistical equilibrium  
 covariance, $\C_\infty$, 
 satisfying in matrix form the  Lyapunov equation:
  \begin{equation}
\A \C_\infty + \C_\infty \A^\dagger = - \I~,
\label{eq:cinf}
\end{equation}
where $\A$ is the operator,  in matrix form, governing the linear dynamics associated with Eq. \eqref{eq:STM},  $\A^\dagger$ is
 the hermitian transpose of $\A$, and
$\I$ is the spatial covariance $\langle {\bf f} ~ {\bf f}^\dagger \rangle$ appropriate  for  the white in space
stochastic excitation, $\bf f$,  with the implication that equal energy input is imparted to each degree of freedom 
(cf. \cite{Farrell-Ioannou-1993e,Farrell-Ioannou-1996a}). As mentioned above, this
choice of excitation has the attribute that the forcing does not impart a bias to the structure.

However, in turbulence the mean flow is time-dependent and the time invariant formulation
of the STM producing  the infinite horizon fluctuation covariance $\C_\infty$ is not an appropriate
model for the covariance arising in a time-dependent mean-flow. In DNS and RNL the coherence time for  fluctuation growth 
is limited by the temporal coherence of the mean flow.  Typical  temporal coherence of the mean flow is 
of the order $T_{d} = {\cal O} (1 ~h/u_\tau)$ \citep{Lozano-Duran-etal-2021},  which corresponds in 
our simulation to $T_d = O( 20 ~h/U)$, 
and it is therefore  appropriate  to restrict the fluctuation development
to extend over a time interval,  $T_d$, consistent with this coherence time. 
In relating the POD mode structure to the STM this coherence time is appropriate for both DNS 
and the RNL. However, while in the  DNS the excitation term $\bf f$ has traditionally been related to the 
fluctuation-fluctuation nonlinearity,
in the case of the RNL a similar excitation  arises from  the 
effective nonlinear scattering of the fluctuations by their  interaction
with the time-dependent mean flow.

It can be shown (cf.  \citep{Farrell-Ioannou-1998a}) 
that the  covariance   of the dynamics governed by Eq. \eqref{eq:STM}  when restricted in time to $T_d$
is 
\begin{eqnarray}
\C_{T_d} &= &e^{\A T_d} \I ~e^{\A^\dagger T_d}+\int_0^{T_d}  ds~e^{\A s} \I ~e^{\A^\dagger s} \nonumber \\
&=&  e^{\A T_d} e^{\A^\dagger T_d} +
\C_\infty - e^{\A T_d} \C_\infty e^{\A^\dagger T_d}~.
\label{eq:T}
\end{eqnarray} 

An alternative to limiting the temporal extent over which fluctuations develop is the inclusion in Eq. \eqref{eq:STM}
of an appropriate eddy viscosity
\citep{DelAlamo-Jimenez-2006}.
Either intervention in the linear dynamics of Eq. \eqref{eq:STM}
has been shown  to result in dynamical structures and spectra commensurate with those observed
\citep{Butler-Farrell-1993,Farrell-Ioannou-1998a,DelAlamo-Jimenez-2006,Hwang-Cossu-2010a,Marusic-etal-2019}.

 Eigenalysis of the covariance $\C_{T_d}$ determines the POD modes as predicted by the STM.  This   covariance is obtained by integration of  \eqref{eq:T}
 in which the initial state covariance, $\I$, is evolved over a time $T_d$ while also being continuously excited  with  covariance $\I$. 
We have obtained best agreement with the observed POD modes in both  DNS and RNL 
when we choose  a disruption time of  $T_d=30$ in the STM. 
The covariance obtained
from Eq. \eqref{eq:T} reflects both the influence of the transient growth of an unbiased initial state as well as the 
accumulated transient growth of an unbiased excitation over the coherence interval $T_d$.
We find that in this problem for $T_d=30$ the initial condition does not appreciably influence the covariance.

Because the streamwise mean streak is mirror symmetric in the
spanwise direction,   the POD modes of the STM   will be either
 sinuous or varicose.  We find that in  both DNS and RNL, as well as in the STM, the top POD modes are
 sinuous.  This result is consistent with the optimally growing  perturbation to a low-speed streak being of sinuous form.
It is worth noting the further consistency in the coincidence of the sinuous optimal with the low speed streak arising from
the fact that  the Reynolds stresses of the sinuous fluctuations  are  favorably configured
to amplify the low-speed streak through the lift-up process \citep{Farrell-Ioannou-2022,Nikolaidis-RS-2023}. 

The top  POD modes with  streamwise wavenumber $n_x=1,2,3$ of the STM are shown in Figures
\ref{fig:dns_stm_1}, \ref{fig:dns_stm_2} and \ref{fig:dns_stm_3} next to the corresponding POD modes of
the DNS. The STM POD modes obtained from the RNL mean flow are similar to those obtained from  the DNS mean flow 
and are not shown.
This similarity in structure is expected because the mean low-speed streaks in DNS and RNL have similar structure,
 as seen in Figs. \ref{fig:ali2} and \ref{fig:FC}.

The dominant POD modes of the STM (cf. Figures   \ref{fig:dns_stm_1}, \ref{fig:dns_stm_2} and \ref{fig:dns_stm_3}) exhibit  complex three dimensional 
velocity fields with striking similarity  in  structure and phasing 
to those of DNS and RNL  (cf. Figures \ref{fig:dns_k1s}, \ref{fig:dns_k2s} and \ref{fig:dns_k3s}) indicative of a parallel  mechanism  underlying  their dynamics.
This  similarity in the dynamical structure of the POD modes
among STM, DNS and RNL argues strongly  for identifying  the dynamical origin of the fluctuation variance in
DNS and RNL with  the growth  of   optimal perturbations  in the mean flow streak.
This identification of the origin of perturbations on the R-S with optimally growing structures 
explains the similarity in
structure of the POD modes   in DNS and RNL 
as consistent with the robust dynamics of optimal perturbation growth in shear flow.

}
\section{Discussion and Conclusion}


POD analysis was carried out on  a DNS of  turbulent Poiseuille flow at $R=1650$
 and the corresponding
quasi-linear RNL simulation.
The RNL system was chosen for this comparison because 
it is dynamically similar to S3T so that the S3T$\rightarrow$RNL $
\rightarrow$DNS sequence of dynamical systems
form a conceptual bridge connecting the analytically comprehensive 
characterization of turbulence in S3T to DNS turbulence, which lacks 
a similarly complete analytic characterization.
The motivation for this work is to 
exploit this conceptual bridge to extend the comprehensive understanding of S3T dynamics to obtain a similar comprehensive understanding  of the dynamics of
DNS.

The POD modes analyzed were chosen to correspond to the 
first and second cumulants of the S3T SSD, these being
the streamwise-mean flow and 
the covariance of fluctuations from it. In general this SSD is closed by parameterizing  the third cumulant 
using  stochastic excitation.  In the present case
this stochastic excitation has been set to zero.
In RNL the equivalent to the covariance  in S3T is  the covariance of  the  Lyapunov vectors with zero Lyapunov exponent of the time-varying 
linear operator linearized about the fluctuating streamwise-mean flow. 
The Lyapunov vectors that are spontaneously emergent in  the dynamics of RNL turbulence are
analogous  to neutral eigenmodes supporting a time independent mean state. 
The structure of the first cumulant 
corresponds to the streamwise-mean dynamics subject to forcing arising from  the  Reynolds stresses of 
the analytically known structures of the second cumulant. 
And finally, the statistical state of the turbulence is regulated by feedback from the second cumulant
to bring the time-varying streamwise-mean flow to neutral stability, 
in the sense that the characteristic Lyapunov exponent of the linear operator governing the second cumulant
is exactly  zero. 
As an illustrative example of the power and utility of being in possession of an analytic theory 
for the dynamics of wall-turbulence consider the problem of understanding the mechanism 
determining the statistical mean state of the turbulence.   Of all the possible mechanisms that one 
might hypothesize, this mechanism is identified analytically in S3T-RNL to be modification of the 
time-dependent streamwise mean state by Reynolds stress feedback arising from the fluctuations
specifically to bring the characteristic Lyapunov exponent of the linear 
fluctuation equation to the real number zero.  Extensive study of DNS data has verified that the 
characteristic exponent of the DNS streamwise mean state corresponds to this parametric 
growth stabilization mechanism \citep{Nikolaidis-etal-Madrid-2018, Nikolaidis-alpha-2022}. It is worth noting that this mechanism of regulating turbulence 
to its statistical mean state  by feedback regulation operating between the fluctuations and the mean state 
 such as to stabilize the mean state to linear instability  is similar to that hypothesis by \cite{Malkus-1956} 
 which posited that the statistical state of  turbulence is determined by feedback
regulation  to neutrality of the mean-state's inflectional modes. This hypothesis  was not verified for the case of wall-bounded turbulence \citep{Reynolds-Tiederman-1967}, but needed as we have seen above only substitution of  parametric neutrality of the   time and spanwise  varying streamwise-mean flow
for the posited inflection mode neutrality of the temporal mean flow
to correspond with the mechanism regulating the statistical mean state of wall-turbulence.


In our study of the POD modes we considered first the streamwise-mean component of the turbulence.
We found close correspondence in structure  between  the POD modes  of  the DNS and RNL fields. 
An initial interpretation of this similarity suggested  that the scale invariant R-S
formation mechanism analytically identified in the S3T-RNL SSD is also operating in DNS.  
Although scale invariant R-S dynamics provides a possible explanation for the POD 
modes found in both DNS and RNL, the random phase assumption, which is traditionally   taken to characterize the POD 
modes in directions  in which  solutions to the equations are statistically homogeneous,  is not necessarily 
valid when a mechanism of symmetry breaking is active.
In the present case an   instability process, 
 which has analytic expression in the SSD stability analysis of the stability 
 of streamwise and spanwise homogeneous mean flows, occurs to break the spanwise symmetry, resulting in R-S structures \citep{Farrell-Ioannou-2012,Farrell-Ioannou-2017-bifur}.
If this process is active it would imply that the random phase assumption in the spanwise for
the POD $k_x=0$ modes is not valid. 
In order to examine this possibility we aligned the most prominent low-speed streak of the flow to obtain a spatially coherent  
time-averaged low speed streak and determined the spanwise Fourier components of this coherent streak. 
We then verified that the Fourier components of this coherent streak corresponded to the structure and the amplitudes of the POD modes
that were obtained making the random phase assumption. 
Furthermore, 
we have verified that 
removing the random phase assumption among the POD
modes of harmonic form by aligning them to zero phase difference revealed that
the POD modes constituted the Fourier components of the coherent 
non-harmonic R-S structure educed by aligning the low-speed streak in the 
turbulence simulation. 

 After having determined the temporal mean structure of the R-S   by collocation, it remained
  to identify the   $k_x\ne0$ 
 POD modes associated with this R-S.
In order to  isolate the POD modes of the $k_x \ne 0$ fluctuations
that are associated with the R-S structure,  the associated fluctuation velocity fields were also  translated
to be aligned  with the collocated low speed streak.
 POD analysis of the aligned velocity fields
 revealed  close correspondence between the $k_x\ne 0$  POD modes in DNS and RNL.  
 The dominant streamwise varying POD modes in both DNS and RNL were found to be characterized by
 a prominent component of 
 streak-localized sinuous oblique waves  that had previously been identified in 
 analysis of the S3T SSD \citep{Farrell-Ioannou-2012}. In those studies  sinuous oblique waves  were shown to give rise to 
 Reynolds stresses properly collocated with the streak to force roll circulation 
 that amplify the streak through the lift-up process.  Moreover, these sinuous oblique waves are exactly 
 the structures that are predicted to arise from a turbulent background flow field because they are 
 the optimally growing perturbations.
  In this paper it  was confirmed using an STM that the sinuous POD modes found in both our DNS and RNL
 similarly correspond closely  to  the time average response to unbiased  stochastic excitation and  is therefore consistently
 dominated by the optimally growing perturbations.
 In a companion paper \citep{Nikolaidis-RS-2023} we show, using the same DNS and RNL dataset,
  that the Reynolds stress of the sinuous POD modes in both DNS and RNL 
 sustain the low-streak streak while the varicose POD modes  suppress it, and that the effect  of the
 sinuous and varicose POD modes is reversed in the case of   high-speed streaks. 

  Although we have connected the POD modes in both DNS and RNL to the growth of optimal perturbations through
  the STM analysis,  we have not addressed in this paper the mechanism by which these optimal perturbations are excited.
  In the case of RNL this is the parametric growth process while in the case of DNS
  it is a combination of the parametric growth process and excitation by the  fluctuation-fluctuation nonlinearity.
  This question is of importance because, to the extent that  parametric growth dominates the fluctuation dynamics in DNS turbulence,
   DNS turbulence inherits  the  analytic characterization of  RNL turbulence. While this paper does not settle this question,
   ongoing work indicates that the parametric growth mechanism does  dominate in DNS \citep{Nikolaidis-etal-Madrid-2018,Nikolaidis-alpha-2022}.
   
%




\subsection*{Acknowledgments}
This work was funded in part by the Second Multiflow Program of the European Research Council. 
Marios-Andreas Nikolaidis gratefully acknowledges the support of the Hellenic Foundation for Research and Innovation (HFRI) and the General Secretariat 
for Research and Technology (GSRT), under the HFRI PhD Fellowship grant 1718/14518. Brian F. Farrell  was partially supported by NSF AGS-1640989. Adri\'an~Lozano-Dur\'an was supported by National Science Foundation under Grant No. 2140775.
We would like to thank Javier Jim\'enez, Dennice Gayme, Daniel Chung, Georgios Rigas and Navid Constantinou and the three anonymous reviewers for their helpful  comments. 

 \section*{Declaration of interest}

The authors report no conflict of interest

\appendix

\section{Construction of  the covariances with symmetry restrictions}\label{sec:symmetries}

\begin{figure*}
\centering
\includegraphics[width=30pc]{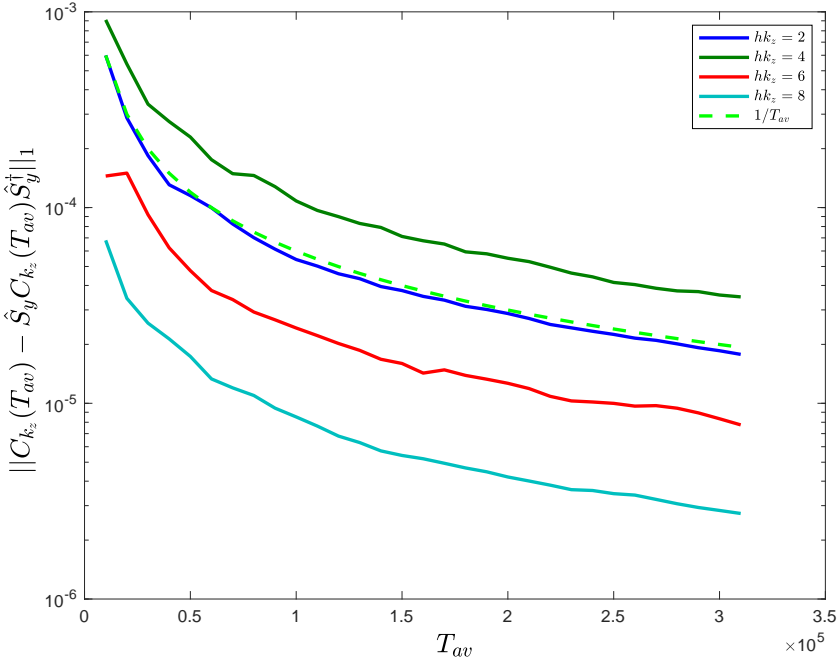}
\vspace{-1em}
\caption{The 1-norm of the difference $C_{k_z} - \hat{S}_y C_{k_z}\hat{S}_y^\dagger$ between the covariance matrix  $C_{k_z}$  \eqref{eq:Ckz},
and the covariance of the reflected flow about the $x-z$ plane 
at 
the  center  of the flow ($y=1$)  as a function of the averaging time, $T_{av}$, for  $h k_z=2,4,6,8$ for a DNS of NL100.
$\hat{S}_y$ is defined in Appendix A,  Eq. \eqref {eq:B5}.
This plot verifies that  reflection symmetry about the centerline  is a statistical symmetry of the flow and that this symmetry
is approached at the rate $1/T_{av}$ consistent with the law of large numbers for quadratic statistics. Time is non-dimensionalized by $h/U$.} 
\label{fig:P}
\end{figure*}

Homogeneity in the streamwise and spanwise directions allows the decomposition of velocity field snapshots into sums of plane waves with Fourier coefficients that depend on the wall-normal direction. Application of mirror symmetries in $y$ and $z$ incorporates the 2-point statistics from the total flow field into a single covariance for each $\vert k_x \vert, \vert k_z \vert$ wavenumber pair. Convergence  towards these statistical symmetries  is slow.
For example in Fig. \ref{fig:P} we demonstrate the slow convergence of the statistics to 
the asymptotic mirror symmetric state about the wall-normal plane at the center of the channel.

For a single $k_z,k_x$ pair the three components of the  velocity field is comprised by two independent plane waves
 

 \begin{eqnarray}
\mathbf{\Phi}_{k_x}e^{ik_x x}= \left(
\begin{array}{c}
  ~A_{k_x,k_z}(y)\\
   ~B_{k_x,k_z} (y)\\
   ~\Gamma_{k_x,k_z}(y)
\end{array}
\right)
e^{i(k_x x + k_z z)}
+ \left(
\begin{array}{c}
  ~A_{k_x,-k_z}(y)\\
   ~B_{k_x,-k_z} (y)\\
   ~\Gamma_{k_x,-k_z}(y)
\end{array}
\right)
e^{i(k_x x - k_z z)} ~.
\end{eqnarray}
With $A$ we denote the streamwise component of the velocity field, $B$ the wall-normal and with $\Gamma$ the spanwise component.
A special case is the $k_x=0$ component for which the coefficients of $k_z$ and $-k_z$ will be complex conjugates.
The two symmetries we consider are mirror symmetry in $y$ with respect to the half-channel $x-z$ plane at $y=1$ and in $z$ with respect to the half width plane $x-y$ at $z=\pi/2$. Those produce a fourfold increase in the amount of data that will be included in the covariance matrix. 


First we consider the spanwise mirroring operation. This will transform $z$ to $\pi-z$ and change sign in the spanwise velocity component.

 \begin{eqnarray}
\hat{S}_{z} \mathbf{\Phi}_{k_x}e^{ik_x x}=  \left(
\begin{array}{c}
  A_{k_x,-k_z}(y)\\
  B_{k_x,-k_z} (y)\\
  -\Gamma_{k_x,-k_z}(y)
\end{array}
\right)
e^{i (k_x x + k_z (z -\pi))}
+ \left(
\begin{array}{c}
  A_{k_x,k_z}(y)\\
  B_{k_x,k_z} (y)\\
  -\Gamma_{k_x,k_z}(y)
\end{array}
\right)
e^{i (k_x x - k_z (z - \pi))} ~ 
\end{eqnarray}
The $-i k_z \pi$ phase that appears in the plane wave will cancel out when the covariance is formed.

In the wall-normal mirroring the effect is to transform $y$ to $2-y$ and change sign in the wall-normal velocity component,

\begin{eqnarray}
\hat{S}_{y} \mathbf{\Phi}_{k_x}e^{ik_x x}  &= & \left(
\begin{array}{c}
  A_{k_x,k_z}(2-y)\\
   -B_{k_x,k_z} (2-y)\\
   \Gamma_{k_x,k_z}(2-y)
\end{array}
\right)
e^{i (k_x x + k_z z )}
+ \left(
\begin{array}{c}
  ~~A_{k_x,-k_z}(2-y)\\
   ~~-B_{k_x,-k_z} (2-y)\\
   ~~\Gamma_{k_x,-k_z}(2-y)
\end{array}
\right)
e^{i (k_x x - k_z  z)} ~  \nonumber
\\
\end{eqnarray}
What the $2-y$ coordinate implies is that the wall-normal structure will be inverted for each component.
Summarizing the above operations, the total covariance will be comprised by the individual covariances obtained for each of the 4 components below 


\begin{eqnarray}
\Phi_{k_z} = \left(
\begin{array}{c}
  ~~A_{k_z}(y)\\
  ~~B_{k_z} (y)\\
   ~~\Gamma_{k_z}(y)
\end{array}
\right)
e^{i k_z z}~,
\hat{S}_{z} \Phi_{k_z} = \left(
\begin{array}{c}
  ~~A_{-k_z}(y)\\
   ~~B_{-k_z} (y)\\
   -\Gamma_{-k_z}(y)
\end{array}
\right)
e^{i k_z z}~,\notag
\end{eqnarray}
\begin{eqnarray}
\hat{S}_{y} \Phi_{k_z} = \left(
\begin{array}{c}
  ~~A_{k_z}(2-y)\\
   -B_{k_z} (2-y)\\
   ~~\Gamma_{k_z}(2-y)
\end{array}
\right)
e^{i k_z z}~,
\hat{S}_{z}\hat{S}_{y} \Phi_{k_z} = \left(
\begin{array}{c}
  ~~A_{-k_z}(2-y)\\
   -B_{-k_z} (2-y)\\
   -\Gamma_{-k_z}(2-y)
\end{array}
\right)
e^{i k_z z}~,
\end{eqnarray}
where the $k_x$ subscript has been omitted.

We form the covariance obtained from the initial wave. To highlight the inner structure of this covariance due to the different velocity components the following representation is chosen 


\begin{eqnarray}
C_{k_z} = \left(
\begin{array}{ccc}
  C_{k_z}^{uu} & C_{k_z}^{uv} & C_{k_z}^{uw} \\
  C_{k_z}^{vu} & C_{k_z}^{vv} & C_{k_z}^{vw} \\
  C_{k_z}^{wu} & C_{k_z}^{wv} & C_{k_z}^{ww}
\end{array}
\right)~,
\end{eqnarray}
with $C_{k_z}^{u_iu_j}=(C_{k_z}^{u_ju_i})^\dagger$. In the following the $k_z$ subscript will be omitted where possible  and instead of $u_iu_j$ the superscript $ij$ will be used. 
So the covariance can be written as:
\begin{eqnarray}
C = \left(
\begin{array}{ccc}
  C^{11} & C^{12} & C^{13} \\
  C^{21} & C^{22} & C^{23} \\
  C^{31} & C^{32} & C^{33}
\end{array}
\right)~.
\end{eqnarray}

Statistical symmetry in reflections of the velocities in $z$ merge the covariance of the $-k_z$ component with the $k_z$. The negative $k_z$ covariance will be modified to account for this symmetry



\begin{eqnarray}
\hat{S}_zC_{-k_z}\hat{S}_z^\dagger = \left(
\begin{array}{ccc}
  ~~(C^{11}) & ~~(C^{12}) & -(C^{13}) \\
  ~~(C^{21}) & ~~(C^{22}) & -(C^{23}) \\
  -(C^{31}) & -(C^{32}) & ~~(C^{33})
\end{array}
\right)~.
\end{eqnarray}
Reflections in $y$ require to reverse the order of the row and column indexes in each individual covariance  and  if this operation is noted as $\hat{S}_y C^{ij}\hat{S}_y^\dagger = C^{ij}_R$
we have:
\begin{eqnarray}
\hat{S}_yC\hat{S}_y^\dagger = \left(
\begin{array}{ccc}
  ~~C^{11}_R & -C^{12}_R & ~~C^{13}_R \\
  -C^{21}_R & ~~C^{22}_R & -C^{23}_R \\
 ~~C^{31}_R & -C^{32}_R & ~~C^{33}_R
\end{array}
\right)~.\label{eq:B5}
\end{eqnarray}


The total covariance will be comprised by the following components

\begin{eqnarray}
C^{t}_{k_z}=(C_{k_z}+\hat{S}_yC_{k_z} \hat{S}_y^\dagger+ \hat{S}_zC_{-k_z}\hat{S}_z^\dagger + \hat{S}_y\hat{S}_zC_{-k_z} \hat{S}_z^\dagger\hat{S}_y^\dagger
\label{eq:B9})/4
\end{eqnarray}

To account correctly for the relative energy between $k_z=0$ and $k_z \ne 0$ components the eigenvalues of covariances with $k_z \ne 0$ are doubled in the ordering process.

\bibliographystyle{jfm}
\end{document}